\documentclass[preprint2]{aastex}

\begin{document}


\title{On Potassium and Other Abundance Anomalies of Red Giants in NGC 2419}


\author{C. Iliadis\altaffilmark{a}}
\affil{Department of Physics \& Astronomy, University of North Carolina, Chapel Hill, NC 27599-3255}
\affil{Triangle Universities Nuclear Laboratory, Durham, NC 27708-0308}
\email{iliadis@unc.edu}

\author{A. I. Karakas}
\affil{Research School of Astronomy \& Astrophysics, Australian National University, Canberra ACT 2611, Australia}

\author{N. Prantzos}
\affil{Institut d'Astrophysique de Paris, UMR7095 CNRS, Universit\'e P. \& M. Curie, 98bis Bd. Arago, 75104 Paris, France}

\author{J. C. Lattanzio}
\affil{Monash Centre for Astrophysics, School of Physics \& Astronomy, Monash University, Victoria 3800, Australia}


\author{C. L. Doherty}
\affil{Monash Centre for Astrophysics, School of Physics \& Astronomy, Monash University, Victoria 3800, Australia}



\altaffiltext{a}{Visiting Scientist, Monash Centre for Astrophysics, School of Physics and Astronomy, Monash University, Victoria 3800, Australia.}


\begin{abstract}
Globular clusters are of paramount importance for testing theories of stellar evolution and early galaxy formation. Strong evidence for multiple populations of stars in globular clusters derives from observed abundance anomalies. A puzzling example is the recently detected Mg-K anticorrelation in NGC 2419. We perform Monte Carlo nuclear reaction network calculations to constrain the temperature-density conditions that gave rise to the elemental abundances observed in this elusive cluster. We find a correlation between stellar temperature and density values that provide a satisfactory match between simulated and observed abundances in NGC 2419 for all relevant elements (Mg, Si, K, Ca, Sc, Ti, and V). Except at the highest densities ($\rho \gtrsim 10^8$~g/cm$^3$), the acceptable conditions range from $\approx$ $100$~MK at $\approx$ $10^8$~g/cm$^3$ to $\approx$ $200$~MK at $\approx$ $10^{-4}$~g/cm$^3$. This result accounts for uncertainties in nuclear reaction rates and variations in the assumed initial composition. We review hydrogen burning sites and find that low-mass stars, AGB stars, massive stars, or supermassive stars cannot account for the observed abundance anomalies in NGC 2419. Super-AGB stars could be viable candidates for the polluter stars if stellar model parameters can be fine-tuned to produce higher temperatures. Novae, either involving CO or ONe white dwarfs, could be interesting polluter candidates, but a current lack of low-metallicity nova models precludes firmer conclusions. 
We also discuss if additional constraints for the first-generation polluters can be obtained by future measurements of oxygen, or by evolving models of second-generation low-mass stars with a non-canonical initial composition.
\end{abstract}




\section{Introduction}\label{sec:intro}
Globular clusters represent fascinating puzzles, particularly after it was discovered that they consist of multiple populations of stars. Distinct populations within a given globular cluster manifest themselves by discrete tracks in the color-magnitude diagram \citep[][]{villanova07,piotto07} and a negative correlation (anticorrelation) of abundances between pairs of light elements, such as C-N, O-Na, and Mg-Al. For a recent review, see \citet[][]{gratton12}. Since the anticorrelations are mostly absent\footnote{A small fraction of field stars exhibit abundance anomalies characteristic of some globular cluster stars. These field stars are believed to have escaped during the dynamical history of their parent globular cluster \citep[see][and references therein]{lind15}.} in halo field stars \citep[][]{gratton00}, their origin must be related to the poorly understood processes of globular cluster and early galaxy formation. 

The abundance anticorrelations have been detected both in red giants and in unevolved stars \citep[][]{gratton01}. Since these low-mass stars cannot produce the high temperatures required to alter the abundances of O, Na, Mg, or Al \citep[][]{powell99,gratton04}, the reported abundance anomalies likely originate from an earlier (i.e., first-generation globular cluster) stellar population that subsequently polluted the matter out of which the currently observed second-generation globular cluster stars formed\footnote{For a different interpretation that does not involve multiple stellar {\it generations}, see \citet[][]{bastian13}.}. The latter stars presumably formed in a dense environment occupied by the first-generation stars. The observations confront us with a number of crucial questions. How many star-forming episodes took place in globular clusters? How does the process of star formation differ in a dense cluster environment compared to a molecular cloud devoid of stars? What kind of first-generation stars gave rise to the abundance anticorrelations and what was their composition? 

The measured abundance anticorrelations differ in magnitude from cluster to cluster, even among the clusters that show no significant spread in the iron content. The Mg-Al anticorrelation, for example, is not observed in some globular clusters. This suggests that the pollution mechanism and the nature of the polluter stars (also called {\it polluters}) may vary, depending on the total mass and metallicity of the cluster \citep[][]{carretta09,meszaros15}. 

For the cluster NGC 6752, the measured stellar abundance anomalies involving O, Na, Mg, and Al can be explained by hydrogen burning at moderate temperatures, near $75$~MK, as was shown by \citet[][]{prantzos07}. Candidate sources for the first-generation polluter stars include rapidly rotating massive stars \citep[][]{decressin07}, massive stars in interacting binary systems \citep[][]{demink09}, stellar collisions \citep[][]{sills10},  supermassive stars with $M$ $\approx$ $10^4~M_\odot$ \citep[][]{denissenkov14}, intermediate-mass asymptotic giant branch (AGB) stars \citep[][]{dantona02}, super-asymptotic giant branch stars (SAGB) \citep[][]{ventura12}, and novae \citep[][]{smith96,maccarone12}, but detailed stellar models fail to account for all  observations. 

The recent discovery of a Mg-K anticorrelation among red giant stars in the cluster NGC 2419 \citep[][]{mucciarelli12,cohenkirby12} adds to the mystery of globular cluster abundance anomalies. The strong potassium enhancements, correlated with magnesium depletions, cannot be explained by hydrogen burning at moderate temperatures, near $75$~MK, where the Coulomb barrier gives rise to insufficiently small thermonuclear rates of the relevant proton capture reactions. Elevated temperatures will be required to account for the reported potassium enhancements. Recently, potassium enhancements have also been measured in NGC 2808 \citep{mucciarelli15}, albeit of a much smaller magnitude.

The nature and origin of the metal-poor cluster NGC 2419 are not yet understood. It is located in the outer halo, further away than the Small and Large Magellanic clouds, at a galactocentric distance of $87.5$~kpc. It is $12.3 \pm 1.3$~Gy old \citep[][]{forbes10}, and has an orbital period of about $3$~Gy \citep[][]{dicris11}. It is the third most massive cluster \citep[$9.12\times10^5$~M$_{\odot}$;][]{ibata11a} in our Galaxy. For a massive globular cluster, it also has an unusually large half-light radius of $24$~pc \citep[][]{ibata11b}. Therefore, it was suggested that NGC 2419 may not be a genuine globular cluster, but rather the remnant of an accreted dwarf galaxy \citep{mackey05}. The recently observed strong potassium enhancements  contribute to the puzzle surrounding this stellar aggregate.

A first attempt to explain the measured strong potassium abundance enhancements that are correlated with large magnesium depletions in the atmospheres of red giants in NGC 2419 was made by \citet[][]{ventura12}. They assumed that the anomalous abundances are produced during hot-bottom burning in massive AGB stars and super-AGB stars, involving temperatures near $150$~MK, and that the second-generation stars formed directly from the ejecta of the first-generation AGB or super-AGB stars. However, a number of parameters, such as thermonuclear reaction rates and the stellar mass loss rate, had to be fine-tuned in their models to account for the reported Mg-K anticorrelation.

We do not know the nature of the first generation polluter stars that gave rise to the reported abundance anomalies in the presently observed second generation. In this work, we will provide a fresh look at this puzzling situation. We choose not to limit ourselves to specific stellar models but will perform nuclear reaction network calculations at constant temperature, $T$, density, $\rho$, amount of consumed hydrogen, $\Delta X_H$, and initial composition, $X_i$, following broadly the ideas presented in \citet{prantzos07}. These parameters are varied in a Monte Carlo procedure to determine the conditions that best reproduce not only the reported Mg-K anticorrelation but {\it all relevant observed abundances} in NGC 2419. This information will be important for identifying the astrophysical sites that gave rise to the puzzling abundance anomalies. The results could have significant implications for models of both stellar and globular cluster formation.

In Section \ref{sec:setup} we discuss our procedure in more detail, including the observations, our nuclear reaction network, the assumed initial composition, and the nucleosynthesis during hydrogen burning. Results are presented in Section \ref{sec:results}. Candidate sources for first-generation polluter stars are discussed in Section \ref{sec:polluters}. In Section~\ref{sec:future}, we comment on possibilities to further constrain the conditions in the polluter candidates. A summary and conclusions are given in Section \ref{sec:summary}.

\section{Procedure}\label{sec:setup}
\subsection{General considerations}\label{subsec:general}
\begin{figure*}
\epsscale{2.0}
\plotone{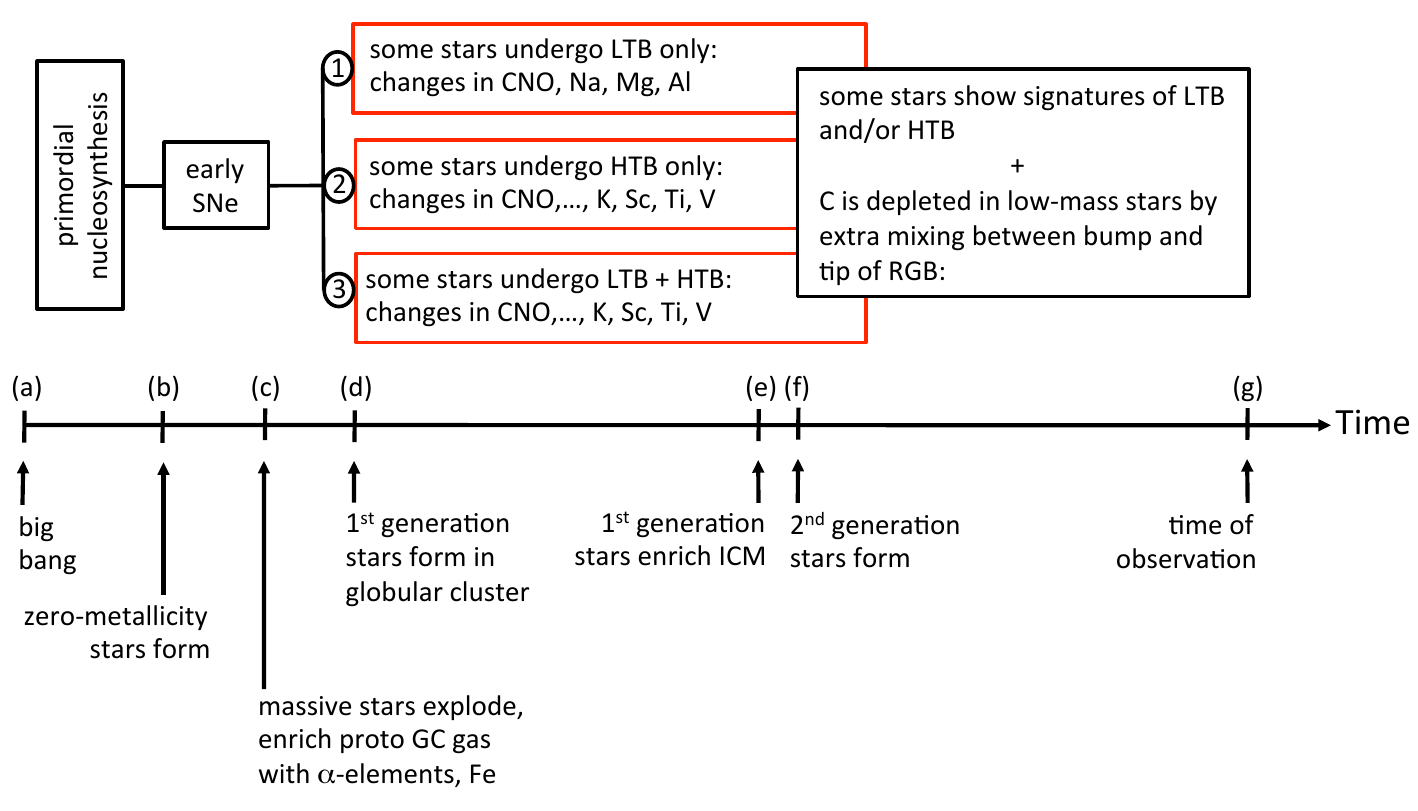}
\caption{Schematic representation of globular cluster evolution: (a) big bang; (b) zero-metallicity stars form; (c) type II supernovae enrich the proto-cluster gas; (d) birth of first-generation globular cluster stars: some may undergo low-temperature hydrogen burning (LTB) only (1), high-temperature hydrogen burning (HTB) only (2), or undergo both processes at different times (3); (e) pollution of intracluster medium (ICM) by first-generation stars; (f) second-generation stars form and evolve until present time (g). The time axis is not to scale: the time between first and second generation star formation, (d) -- (f), is $\lesssim$ $1.0$~Gy, whereas the age of an old cluster, (a) -- (g), is $>$ $10$~Gy. \label{fig:fig1}}
\end{figure*}

The scenarios of globular cluster formation and evolution are schematically shown in Figure~\ref{fig:fig1}. After primordial nucleosynthesis (a-b), the first stars form (b). The ejecta and winds of these zero-metallicity stars, including the contributions from massive stars of subsequent generations, enrich the proto-cluster gas with metals (c). Eventually, the first-generation globular cluster stars form (d). The massive stars among these evolve quickly and explode as type II supernovae. Other first-generation stars (1) undergo hydrogen burning during their evolution, giving rise to the ubiquitous O-Na anticorrelation that we observe in second-generation stars. We will call the process responsible for the anomalous CNO, Na, Mg, and Al abundance pattern {\it low-temperature hydrogen burning} (LTB). Other first-generation stars (2), or the same ones (3), undergo hydrogen burning at higher temperatures, giving rise to the recently discovered Mg-K anticorrelation. We refer to this process as {\it high-temperature hydrogen burning} (HTB). During their evolution, the first-generation stars eject part of their matter (e). The currently observed second-generation stars form from the polluted intracluster gas (f), thereby inheriting the nucleosynthesis signatures of both the pre-enrichment before cluster formation and of the first-generation stars \citep[see, e.g.,][]{dercole08,decressin08}. Today (g), globular clusters consist mainly of old low-mass stars and very little cold gas. 

The above picture is broadly supported by observations. For most globular clusters, the Fe-group elements (Fe, Cu, Ni), $\alpha$-elements (e.g., Si, Ca), s-process abundances (e.g., Ba, Sr), and r-process abundances (e.g., Eu) in a given cluster scatter little from star to star and are independent of the evolutionary phase of the observed globular cluster stars \citep{james04}. This indicates that the globular cluster formed from pre-enriched homogeneous matter. Furthermore, the abundances of the light metals (i.e., C to K) are not correlated with those of the heavy metals (i.e., iron peak, s-, and r-elements), indicating that the site responsible for the light-element abundance anomalies did not produce significant amounts of Fe, s-process, and r-process elements \citep{james04,prantzos06}. This argues strongly against type II supernovae as the first-generation polluters. 

An important step in identifying the polluters is to constrain the physical conditions for hydrogen burning that gave rise to the abundance anomalies. For the well-studied cluster NGC 6752, \citet{prantzos07} applied a simple method that did not depend on any specific stellar model. The method has the additional advantage of making robust predictions once the necessary thermonuclear reaction rates are reliably known. They performed a series of hydrogen burning reaction network simulations at constant temperature and density, for a given amount of consumed hydrogen, and identified the conditions that simultaneously reproduced all of the measured light-element abundances, from C to Al. Each network simulation started from an initial composition obtained from observations of field stars with a similar iron content to NGC 6752. It was found that a narrow temperature range around $75$~MK could account for the light-element anomalies, if the nuclearly processed matter is mixed with at least 30\% of pristine (i.e., unprocessed) matter. The reported anomalies, including the ubiquitous O-Na anticorrelation, are then reproduced assuming a range of mixing (or {\it dilution}) factors in excess of $30$\%. They also suggested stellar sites for this low-temperature hydrogen burning, including AGB stars and fast rotating massive stars, but detailed stellar models for both sites have difficulties accounting for the observations \citep{gratton12}. We will apply a similar model to investigate the abundance pattern of light elements measured in NGC 2419 \citep[][]{mucciarelli12,cohenkirby12}. 

\subsection{Observations in NGC 2419}\label{subsec:observations}
\begin{figure*}
\epsscale{2.0}
\plotone{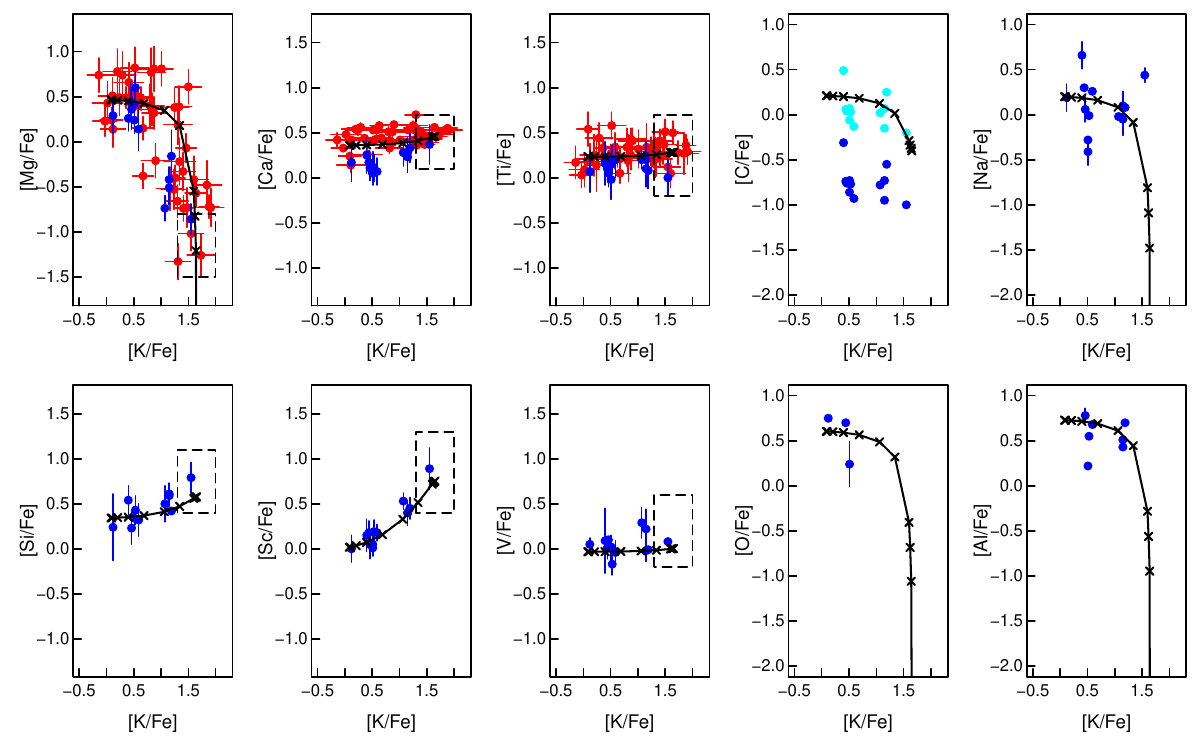}
\caption{Elemental abundances, with respect to Fe, versus K abundance for red giants in NGC 2419; (red) observations of \citet{mucciarelli12}; (dark blue) observations of \citet{cohenkirby12}; (light blue) from \citet{cohenkirby12}, but 0.8~dex was added to [C/Fe] (see text). Simulations are shown in black for the conditions $T$ = $160$~MK, $\rho$ = $900$~g/cm$^3$, and $X_H^f$ = $0.70$: the solid lines are obtained by mixing one part of processed matter with $f$ parts of pristine matter, according to Eq.~(\ref{eq:dilution}); the crosses on the solid lines denote, from left to right, the abundances obtained with dilution factors of $f$ = $1000$, $100$, $30$, $10$, $3$, $1.0$, $0.1$, $0.05$, $0.02$, and $0.0$. In some panels, the cross for $f$ $=$ $0.0$ is off scale. The pristine matter composition (on the left-hand side) is fixed by the initial abundances listed in Table~\ref{tab:composition}, while the (undiluted) processed matter composition (on the right-hand side) is given by the output of the reaction network calculation. The range of acceptable {\it processed} abundances is indicated by the dashed boxes. \label{fig:abund1}}
\end{figure*}

The stars in NGC 2419 have an average metallicity of [Fe/H] $=$ $-2.09 \pm 0.02$\footnote{According to common convention, abundances are given as $\left[ A/B \right]_\star \equiv \mathrm{log}_{10}(N_A/N_B)_\star -  \mathrm{log}_{10}(N_A/N_B)_\odot$, where $N_i$ are number abundances of elements $A$ and $B$ observed in a star ($\star$) or the Sun ($\odot$); while the quantity $\left[ A/B \right]$ is unitless, differences between two values are expressed in units of $dex$ (``decimal exponent").}, with no intrinsic spread in the iron abundance \citep{mucciarelli12,cohenkirby12}. Measured elemental abundances versus potassium abundance for red giants in NGC 2419 are shown as data points in Figure~\ref{fig:abund1}. The red and blue data points are adopted from \citet{mucciarelli12} and \citet{cohenkirby12}, respectively. All abundances were derived from a local thermodynamic equilibrium (LTE) analysis using one-dimensional model atmospheres. The sole exception is the potassium abundance reported by \citet{mucciarelli12}, who applied a unique value of $-0.3$~dex for the non-LTE correction to all targets. We decided to adopt all reported abundances at face value, as was done in other work \citep{ventura12,bellazzini13}. Attempting to place the two studies onto the same scale is not trivial, given their differences in spectral resolution, analysis technique, application of uniform NLTE corrections, and adopted solar abundance. 

The first panel shows the Mg-K anticorrelation: when the K abundance is low, the Mg abundance is high (Mg-normal); when K is strongly enhanced (by $\approx$ $1.5$~dex), Mg is strongly depleted (by $\approx$ $2.0$~dex). Notice that about 30\% of the observed red giants show strong potassium enhancements. {\it In other words, about 30\% of the stars in NGC 2419 consist of matter that underwent an unknown process.}

Both Si and Sc show an increase in abundance for the K-enhanced stars, by about 0.5 dex. The abundances of Ca, Ti, and V are approximately constant for a wide range of K abundances. The panels on the right show the abundances of elements in the CNO and Na--Al region: Na and Al reveal no correlation with K, but a large scatter instead; O is only reported in three stars. 

Carbon (fourth panel in top row) is the only element shown in Figure~\ref{fig:abund1} whose abundance is affected by the evolution of the second-generation stars we currently observe. Because of the distance of NGC 2419, the red giants with measured carbon are all located at the tip of the red giant branch \citep[see Figure 1 in][]{cohenkirby12}. If all these observed stars were born with a composition typical for field stars of the same metallicity, they would have depleted C during their evolution from the luminosity function bump to the tip of the red giant branch. For the cluster NGC 2419, with an average metallicity of [Fe/H] =  $-2.1$, the estimated depletion is $\approx$ $0.8$~dex \citep{angelou12}. The expected natal carbon abundance of the currently observed stars, with $0.8$~dex added to the measured values, is shown as light blue data points in Figure~\ref{fig:abund1} (fourth panel in top row). We will discuss this assumption further in Section~\ref{subsec:procedure}. 


The helium abundance has also been inferred in NGC 2419, both for red giants \citep[][]{lee13} and for horizontal branch stars \citep[][]{dicris15}. We will discuss these observations in Section~\ref{subsec:helium}.

The second-generation red giant stars that we observe in NGC 2419 are about $12$~Gy old and, consequently, must have formed very early during globular cluster evolution. The time available for the first-generation stars to pollute the intracluster gas out of which the second-generation stars formed is very uncertain, but is probably less than a few hundred million years.

\subsection{Strategy}\label{subsec:strategy}
Reaction rates for charged particles are highly sensitive to temperature and density. Low-temperature hydrogen burning ($T \approx 75$~MK) explains both the O-Na anticorrelation and the Mg-Al anticorrelation in Galactic globular clusters \citep[][]{prantzos07}. This temperature range will affect the abundances of the observed elements from C to Al, but not of heavier elements. As we shall see, high-temperature hydrogen burning ($T\gtrsim80$~MK) is necessary to account for the potassium enhancements in NGC 2419. Such elevated temperatures will affect the abundances of the observed elements C, Na, Mg, Al, Si, K, Ca, and Sc. 

The sites of low- and high-temperature hydrogen burning have not been identified yet. We do not know if they operated in different first-generation stars, or in the same stars at different locations, or in the same stars at the same location at different times. We also do not know if low-temperature hydrogen burning took place in NGC 2419. Simultaneous observations of oxygen and sodium are only available for three red giants \citep[][]{cohenkirby12} and thus a correlation between these two elements cannot be established at present in this globular cluster. If low-temperature hydrogen burning did take place in NGC 2419, its abundance signatures may have been modified by high-temperature hydrogen burning. We will return to this point in Section~\ref{sec:future}.

In the present work, we will adopt a simple model that is based on the following assumptions: (i) a single-stage, one-zone (high-temperature) hydrogen burning process in first-generation stars; and (ii) mixing of matter processed by the polluters with pristine, natal matter of the globular cluster. In particular, we would like to determine the conditions of constant temperature, constant density, and the amount of consumed hydrogen required to reproduce the abundance anomalies observed in NGC 2419 and, thereby, constrain the physical conditions of the first-generation polluter stars.

In a reaction network simulation, the amount of consumed fuel (hydrogen) depends on the time duration of the nucleosynthesis: the longer the reaction network runs, the more hydrogen is consumed. In many realistic scenarios, convection continuously carries fresh fuel into the burning region and also dilutes the abundances of the hydrogen-burning products. This effect lengthens considerably the duration of hydrostatic nuclear burning compared to the one-zone process adopted in the present work that disregards convection. Also, for explosive nuclear burning, we cannot expect that our one-zone simulations reproduce realistic burning times. For these reasons, the constraints we obtain on the amount of consumed hydrogen, or, equivalently, the time duration of the nuclear burning, are not meaningful. It is very important, however, to reproduce all the relevant observed abundances {\it for the same amount of consumed hydrogen} \citep[see][]{prantzos07}. Otherwise, no meaningful conclusions can be drawn on temperature and density constraints.

\subsection{Nuclear reaction network}\label{subsec:nuclear}
Our reaction network consists of 213 nuclides, ranging from p, n, $^4$He, to $^{55}$Cr. Thermonuclear reaction rates are adopted from STARLIB\footnote{Available at: \texttt{http://starlib.physics.unc.edu/index.html}.} \citep{sallaska13}. This library has a tabular format that contains reaction rates {\it and} rate probability density functions on a grid of temperatures between $10$~MK and $10$~GK. The probability densities can be used to derive statistically meaningful reaction rate uncertainties at any desired temperature. Most reaction rates important for the present work that are listed in STARLIB, including $^{36}$Ar(p,$\gamma$)$^{37}$K, $^{38}$Ar(p,$\gamma$)$^{39}$K, $^{39}$K(p,$\gamma$)$^{40}$Ca, and $^{40}$Ca(p,$\gamma$)$^{41}$Sc, have been computed using a Monte Carlo method, which randomly samples all {\it experimental} nuclear physics input parameters \citep{longland10}. 
For a few reactions, such as $^{37}$Cl(p,$\gamma$)$^{38}$Ar, $^{37}$Ar(p,$\gamma$)$^{38}$K, and $^{39}$K(p,$\alpha$)$^{36}$Ar, experimental rates are not available yet, and the rates included in STARLIB are adopted from nuclear statistical model calculations using the code TALYS. In such cases, a reaction rate uncertainty factor of 10 is assumed. Most of the important reaction rates for studying hydrogen burning in globular clusters are based on experimental nuclear physics information and provide a reliable foundation for robust predictions.

Stellar weak interaction rates, which depend on both temperature and density, for all species in our network are adopted from \citet{oda94} and, if not listed there, from \citet{fuller82}. The stellar weak decay constants are tabulated at temperatures from $T$ = $10$~MK to $30$~GK, and densities of $\rho Y_e$ = $10-10^{11}$~g/cm$^3$, where $Y_e$ denotes the electron mole fraction. For $\beta^+$- and $\beta^-$-decays, the stellar decay constants should converge to their respective laboratory values at the lowest temperature and density grid point. However, this is not the case for electron captures. When network calculations are performed for densities below $\rho$ = $10$~g/cm$^3$, we use the values tabulated at the lowest density grid point ($10$~g/cm$^3$) for the stellar weak decay constants since it would be inappropriate to adopt the laboratory value under such conditions. The interesting case of $^{37}$Ar is discussed in Section~\ref{sec:flows}. Radioactive nuclides ($^{13}$N, $^{14}$O, $^{15}$O, $^{17}$F, etc.) present at the end of a network calculation were assumed to decay to their stable daughter nuclides.

For $^{26}$Al, we included five species: the ground state ($^{26}$Al$^g$), the isomer ($^{26}$Al$^m$), and three higher lying excited states. Gamma-ray transitions between all these levels are explicitly included in our network calculation and, therefore, no artificial assumption about the equilibration of $^{26}$Al is required \citep[see, e.g.,][]{iliadis11}. The decay constants for the $\gamma$-ray transitions connecting the $^{26}$Al levels are listed in STARLIB \citep[for details, see][]{sallaska13}. It should be noted that we do not take into account the density dependence of the electron capture decays for $^{26}$Al$^g$ $\rightarrow$ $^{26}$Mg and $^{26}$Al$^m$ $\rightarrow$ $^{26}$Mg. The first decay is very slow ($T_{1/2}$ = $7.17 \times 10^5$~yr) and is much slower than the competing $^{26}$Al$^g$(p,$\gamma$)$^{26}$Si reaction. The second decay proceeds mainly via $\beta^+$-decay at densities below $10^7$~g/cm$^3$ \citep{fuller80}. Therefore, this effect is very small for the purposes of the present work. For all stellar weak interaction rates, we assumed a factor of $2$ uncertainty.

\subsection{Abundance flows in the ArK region}\label{sec:flows}
\begin{figure}
\epsscale{0.6}
\plotone{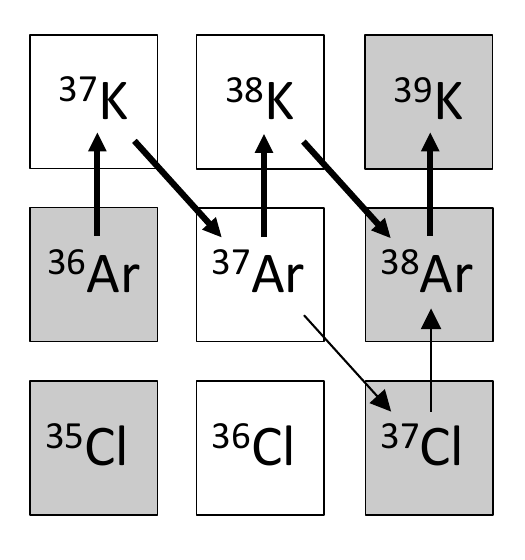}
\caption{Nuclear interactions leading to the synthesis of potassium. Gray boxes indicate stable nuclides. The thick arrows depict the main production channel, while only a minor contribution is expected from the path indicated by the thin arrows. The reason is the significantly faster $^{37}$Ar(p,$\gamma$) reaction compared to the competing electron capture of $^{37}$Ar.\label{fig:flows}}
\end{figure}

We will now consider the interplay of nuclear interactions that gives rise to the synthesis of potassium for the conditions of main interest here. Starting from the most abundant argon isotope, $^{36}$Ar, the main reaction sequence identified by \citet{ventura12} is $^{36}$Ar(p,$\gamma$)$^{37}$K(e$^+$,$\nu$)$^{37}$Cl(p,$\gamma$)$^{38}$Ar(p,$\gamma$)$^{39}$K, and this is repeated by \citet{mucciarelli15}. The notation used by these authors is incorrect, since $^{37}$K can neither capture a positron, nor does it directly decay to $^{37}$Cl. What the authors presumably meant is the reaction sequence $^{36}$Ar(p,$\gamma$)$^{37}$K($\beta^+$$\nu$)$^{37}$Ar(e$^-$,$\nu$)$^{37}$Cl(p,$\gamma$)$^{38}$Ar(p,$\gamma$)$^{39}$K, where $^{37}$K decays via positron emission to $^{37}$Ar, and $^{37}$Ar decays via electron capture to $^{37}$Cl. But this sequence cannot be the main nucleosynthesis path of $^{39}$K either. The situation is depicted in Figure~\ref{fig:flows}. The nuclide $^{37}$Ar decays via electron capture to $^{37}$Cl with a laboratory decay constant of $\lambda_{lab}$ = $2.3 \times 10^{-7}$~s$^{-1}$. Under stellar conditions its decay will depend strongly on the density. For example, at $\rho$ = $10$~g/cm$^3$, the stellar decay constant is $\lambda_{star}$ = $8.5 \times 10^{-10}$~s$^{-1}$ \citep[see also Figure~1.18 in][]{iliadis15}, which is significantly smaller than the decay constant for the competing $^{37}$Ar(p,$\gamma$)$^{38}$K reaction. For increasing density, the electron capture decay constant increases, but the decay constant for the competing (p,$\gamma$) reaction increases as well. In other words, for all conditions of interest here, the main reaction sequence for potassium synthesis is $^{36}$Ar(p,$\gamma$)$^{37}$K($\beta^+$$\nu$)$^{37}$Ar(p,$\gamma$)$^{38}$K($\beta^+$$\nu$)$^{38}$Ar(p,$\gamma$)$^{39}$K, which is indicated by the thick arrows in Figure~\ref{fig:flows}. Only a minor contribution is expected from the branch initiated by the electron capture of $^{37}$Ar, which is depicted by the thin arrows. 

\subsection{Initial composition}\label{subsec:composition}
For the initial composition, we start with the results of a one-zone chemical evolution model for the Milky Way halo, which is an update of \citet[][]{goswami00}. The model reproduces the reported abundances in field stars of the same average metallicity as NGC 2419 (i.e., [Fe/H] = $-2.1$, corresponding to $Z$ = $3.3 \times 10^{-4}$). Our adopted values are listed in Table~\ref{tab:composition}. Similar results can be found in \citet[][]{kobayashi11}. In particular, our initial hydrogen and helium mass fractions amount to $X_H^i$ = $0.754$ and $X_{He}^i$ = $0.245$, respectively. 

The measured abundances of red giants in NGC 2419 (see Section~\ref{subsec:observations}) provide additional information on the initial composition. Specifically, the stars that do not show any K enhancement and Mg depletion are located on the leftmost side in each panel of Figure~\ref{fig:abund1}, near [K/Fe] $\approx$ 0. These stars are not polluted by material that underwent high-temperature hydrogen burning in the previous stellar generation and, therefore, their abundances can be used to constrain the initial composition of the first-generation stars. The abundance values that we matched to the observations in NGC 2419 are shown in boldface in Table~\ref{tab:composition}, and the corresponding original values predicted by the chemical evolution model are listed in the table footnote. The few adjustments are on the level of factors of $2$-$3$. The only exception was $^{27}$Al, whose abundance had to be increased by a factor of $\approx$ $10$ to match the observations. Notice that the halo model of \citet[][]{goswami00} significantly underpredicts the aluminum abundance measured in halo stars at a metallicity of [Fe/H] = $-2.1$ \citep[see their Figure 7 and Figure 4 in][]{andrievsky08}. However, we verified that a variation of the initial $^{27}$Al abundance by an order of magnitude up or down had no impact on our results.

\subsection{Criteria for acceptable solutions}\label{subsec:procedure}
Consider again the measured elemental abundances displayed in Figure~\ref{fig:abund1}. Stars with normal Mg and K abundances are shown on the leftmost side in each panel. Their elemental abundances (i.e., {\it pristine} matter, defined as the composition of the proto-globular cluster gas; Section~\ref{subsec:general}) are in the range of values predicted by Galactic chemical evolution models (Section~\ref{subsec:composition} and Table~\ref{tab:composition}). Stars with the most extreme abundance anomalies are located on the rightmost part in each panel of Figure~\ref{fig:abund1}, near [K/Fe] $\approx$ $2$. The first panel shows the Mg abundance declining by two orders of magnitude with increasing K abundance. Even a small amount of mixing with pristine matter would strongly enhance the Mg abundance and, therefore, the observed extreme values most likely reflect the nearly undiluted composition (i.e., {\it processed} matter) ejected by the polluters. We are seeking the conditions of constant temperature and density that best reproduce these extreme abundance values. Abundances between the pristine and processed matter compositions are obtained in our model by mixing one part of processed matter with $f$ parts of pristine matter. The dilution factors, $f$, are defined by
\begin{equation}
X_{mix} \equiv \frac{X_{proc} + f X_{pris}}{1+f}
\label{eq:dilution}
\end{equation}
where $X_{proc}$ and $X_{pris}$ denote the mass fractions of the reaction network output (i.e., processed matter) and the initial composition (i.e., pristine matter), respectively. We are also seeking the dilution factors that best reproduce the measured extreme abundance values.

The dashed boxes on the right-hand side of some panels in Figure~\ref{fig:abund1} show the ranges of acceptable elemental abundances that we impose on the reaction network output. The boundaries indicated by the dashed boxes are given by   
$1.3$ $<$ [K/Fe] $<$ $2.0$, 
$-1.5$ $<$ [Mg/Fe] $<$ $-0.8$, 
$0.1$ $<$ [Ca/Fe] $<$ $0.7$, 
$-0.2$ $<$ [Ti/Fe] $<$ $0.7$,  
$0.4$ $<$ [Si/Fe] $<$ $1.1$,
$0.4$ $<$ [Sc/Fe] $<$ $1.3$, 
and $-0.2$ $<$ [V/Fe] $<$ $0.6$.
These values are approximations that take into account both the scatter in the data and the abundance uncertainties of individual stars.

For several reasons, we did not impose any boundaries on the carbon abundance. First, carbon may not only take part in {\it hydrogen} burning, but also in other burning episodes of the (first-generation) polluters. For example, in AGB and super-AGB stars of low metallicity ($Z$ $\approx$ $10^{-4}$), carbon is produced during thermal pulses (i.e., helium burning) and destroyed during hot-bottom (i.e., hydrogen) burning during the interpulse period (Section~\ref{subsec:agb}). Whether or not there is a net production of carbon depends on the details of the stellar models \citep{siess10,doherty14}. Second, we already noted in Section~\ref{subsec:observations} that if all of the observed (second-generation) red giants in NGC 2419 were born with a composition typical for field stars of the same metallicity, they would have depleted C during their evolution from the luminosity function bump to the tip of the red giant branch. Correcting the observations (dark blue data points in fourth panel of top row in Figure~\ref{fig:abund1}) for this depletion, we obtained the light blue data points, shown in the same panel. This assumption applies to the K-normal stars (i.e., on the left-hand side in each panel of Figure~\ref{fig:abund1}), but may not be correct for the extreme stars (i.e., on the right-hand side). Since the latter stars were born from matter that underwent an unknown high-temperature hydrogen burning process, their natal abundances are likely different from those of field stars. We will return to this point in Section~\ref{sec:future}.

No other observations of stars in NGC 2419 were used as constraints. In particular, it would be dangerous to impose constraints on Na and Al (panels in last column), since the abundances of these two elements show a significant scatter and no obvious trend with respect to [K/Fe]. As will be seen below, the network calculations that give acceptable solutions within the boundaries listed above will also provide satisfactory fits to the measured Na and Al abundances. The sodium abundance will be discussed in more detail in Section~\ref{subsec:oxna}. 

\section{Results}\label{sec:results}
\subsection{Trial and error solutions}
As already pointed out in Section~\ref{subsec:strategy}, the parameters of our model are: (i) constant temperature, $T$, (ii) constant density, $\rho$, and (iii) final mass fraction of hydrogen, $X_H^f$ = $X_H^i - \Delta X_H$, where $X_H^i$ and $\Delta X_H$ are the initial and consumed hydrogen abundance, respectively.  Besides these parameters, we can also vary the initial composition, $X_i$, and the thermonuclear reaction rates, $N_A\left< \sigma v \right>$. We started with a (fixed) initial composition, given in Table~\ref{tab:composition}, and the (fixed) recommended reaction rates listed in STARLIB. The parameters $T$, $\rho$, and $X_H^f$ were then varied by trial and error to see if an acceptable fit to all measured abundances could be obtained. 

For example, the solid black lines shown in Figure~\ref{fig:abund1} were obtained for the conditions $T$ = $160$~MK, $\rho$ = $900$~g/cm$^3$, and $X_H^f$ = $0.70$. The crosses on the black lines denote, from left to right, the abundances obtained with dilution factors of $f$ = $1000$ (i.e., almost purely pristine matter), $100$, $30$, $10$, $3$, $1.0$, $0.1$, $0.05$, $0.02$, and $0.0$ (i.e., purely processed matter). The simulated processed abundances (on the right-hand side in each panel) satisfy all of the conditions we imposed for Mg, Si, K, Ca, Sc, Ti, and V (indicated by the dashed boxes). At the same time, this particular solution also approximately reproduces the observations for C, O, Na, and Al. Given our best guess of an initial composition and recommended thermonuclear reaction rates, our first main result is that {\it certain combinations of values of constant temperature, density, and consumed hydrogen mass fraction give a satisfactory fit to the abundances of all the relevant elements observed in NGC 2419.} We will next present the results from an automated search. 

\subsection{Monte Carlo sampling: temperature, density, and final H mass fraction}\label{subsec:MC1}
To find sets of parameters that simultaneously satisfy the abundance constraints for Mg, Si, K, Ca, Sc, Ti, and V listed above, we varied the parameters $T$, $\rho$, and $X_H^f$ simultaneously using a reaction network Monte Carlo procedure. At the start of each network calculation, the parameters $\log T$ and $\log \rho$ were randomly sampled according to a uniform probability density (in the ranges of $50$~MK $\le$ $T$ $\le$ $10$~GK and $10^{-4}$~g/cm$^3$ $\le$ $\rho$ $\le$ $10^{11}$~g/cm$^3$). The parameter $X_H^f$ was sampled using a uniform probability density (in the range of  $0.10$ $\le$ $X_H^f$ $\le$ $0.75$). 

\begin{figure}
\epsscale{1.0}
\plotone{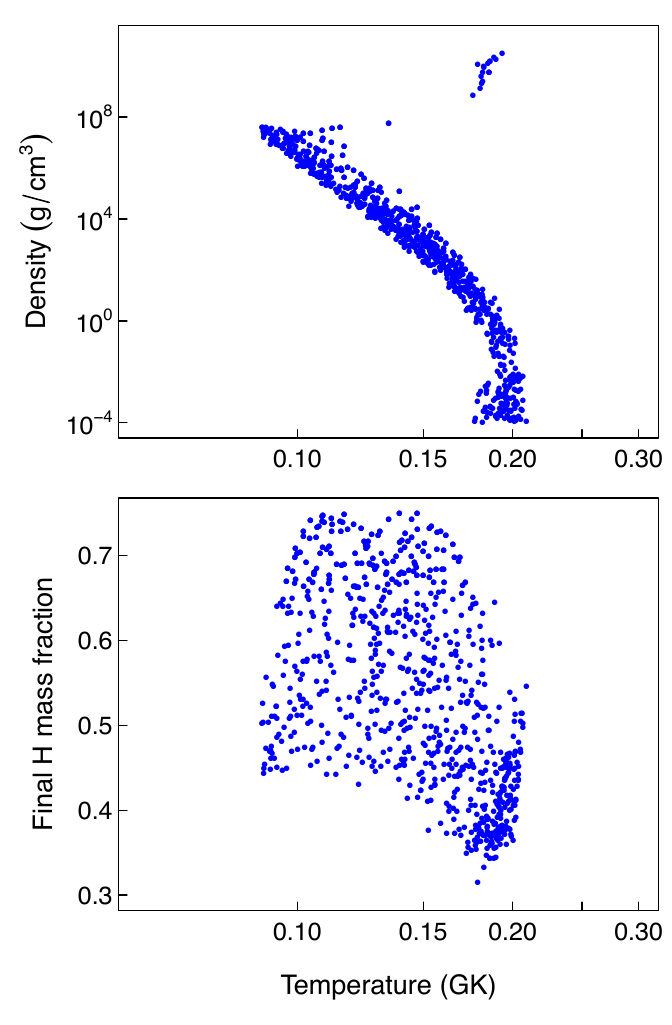}
\caption{(Top) Stellar density versus temperature, and (bottom) final hydrogen mass fraction versus temperature, for sets of ($T$, $\rho$, $X_H^f$) values that reproduce measured elemental abundances in NGC 2419. The results are obtained by random sampling of these three parameters using $5 \times 10^4$ reaction network samples.\label{fig:MC2}} 
\end{figure}

Figure~\ref{fig:MC2} shows part of the sampled ($T$, $\rho$, $X_H^f$) parameter space. Stellar density versus temperature is shown in the top panel, and the final hydrogen mass fraction versus temperature is displayed in the bottom panel. The blue circles show the conditions that simultaneously reproduce the measured abundances of Mg, Si, K, Ca, Sc, Ti, and V. Our second main result is that, given our best guess of an initial composition and recommended thermonuclear reaction rates, {\it we find a correlation between stellar temperature and density values that provide a satisfactory match between simulated and measured elemental abundances in NGC 2419}. Notice that the simulated abundances of Mg, Si, K, Ca, Sc, Ti, and V simultaneously match the observations only for a narrow temperature range of $90$~MK $\le$ $T$ $\le$ $210$~MK. No other temperature conditions, except those indicated by blue circles in Figure~\ref{fig:MC2}, were found in the range between $50$~MK and $10$~GK that reproduced the observed abundances.

For the dilution factors that {\it simultaneosuly} reproduce the most extreme measured abundances (on the right-hand side in each panel of Figure~\ref{fig:abund1}), we find a range of $f$ $=$ $0.01$ $-$ $0.04$. In other words, we do not observe purely processed matter in the stars with the most extreme abundances, although the admixture of pristine matter is very small, consistent with expectation (Section~\ref{subsec:procedure}). This constraint results exclusively from the Mg-K anticorrelation (first panel in Figure~\ref{fig:abund1}). 

The acceptable values of the final hydrogen mass fraction (bottom panel of Figure~\ref{fig:MC2}) scatter over a wide range, i.e., $0.315$ $\le$ $X_H^f$ $\le$ $0.749$. For all solutions shown in Figure~\ref{fig:MC2}, except those at very high densities ($\rho$ $\gtrsim$ $10^7$~g/cm$^3$), the consumed hydrogen mass fraction equals the produced helium mass fraction. We will return to this point in Section~\ref{subsec:helium}, when comparing our simulated helium abundances with recent observations.

The temperature-density correlation shown in the top panel of Figure~\ref{fig:MC2} is interesting. At any given density, simulated and measured abundances can only be matched for a narrow temperature range. The discontinuity at high densities, $\rho$ $\approx$ $10^8$~g/cm$^3$, originates from the onset of electron captures on protons. Outside the region occupied by the blue circles, on the low-$T$ and low-$\rho$ side, too little potassium and too much silicon is produced in the simulations, while silicon is underproduced and calcium is overproduced on the high-$T$ and high-$\rho$ side. In the next section, we will relax our assumptions regarding the initial composition and the nuclear interaction rates.

\subsection{Monte Carlo sampling: initial composition and thermonuclear reaction rates}
The initial mass fractions, $X_i$, of the elements Li, Be, B, N, F, Ne, P, S, Cl, and Ar are not constrained by observations in NGC 2419. So far we adopted for these elements the abundances predicted by a Galactic chemical evolution model that fits abundances of field stars with the same metallicity as NGC 2419 (Section~\ref{subsec:composition} and Table~\ref{tab:composition}). We cannot be certain that our starting abundances correctly predict the initial composition of the polluters. Similarly, we used thus far our best guess for the nuclear interaction rates (i.e., the rates of thermonuclear reactions and weak interactions) provided by STARLIB. But the nuclear rates have uncertainties, either derived from experimental nuclear physics input or from theoretical models (Section~\ref{subsec:nuclear}). 

For these reasons, we repeated the above Monte Carlo procedure for the parameters $T$, $\rho$, and $X_H^f$, but included the initial composition and the nuclear rates in the random sampling. The statistical methods for sampling nuclear interaction rates have been presented recently in the review by \citet{iliadis15b}, and the discussion is not repeated here. It suffices to mention that we adopted a lognormal distribution for both the nuclear rates and the initial abundances, according to
\begin{equation}
f(x) = \frac{1}{\sigma \sqrt{2\pi}} \frac{1}{x} e^{-(\ln x - \mu)^2/(2\sigma^2)}
\end{equation}
where the lognormal parameters $\mu$ and $\sigma$ determine the location and the width, respectively, of the distribution. For a lognormal probability density, samples, $i$, of a nuclear rate or an initial abundance, $y$, are computed from  
\begin{equation}
 \label{eq:sample}
  y_{i} = y_{med}  (f.u.)^{p_{i}}
\end{equation}
where $y_{med}$ and $f.u.$ are the median value and the factor uncertainty, respectively. The quantity $p_{i}$ is a random variable that is normally distributed, i.e., according to a Gaussian distribution with an expectation value of zero and a standard deviation of unity. 

For the nuclear rates, both the median value and the factor uncertainty are provided by STARLIB. We emphasize that the factor uncertainty of experimental Monte Carlo reaction rates depends explicitly on temperature. More information on the adopted nuclear rate factor uncertainties is given in Section~\ref{subsec:nuclear}. For the initial abundances of Li, Be, B, N, F, Ne, P, S, Cl, and Ar, we assumed a factor uncertainty of $f.u.$ = $2.5$ in the absence of more information. All nuclear rates and initial abundances were sampled independently. 

\begin{figure}
\epsscale{1.0}
\plotone{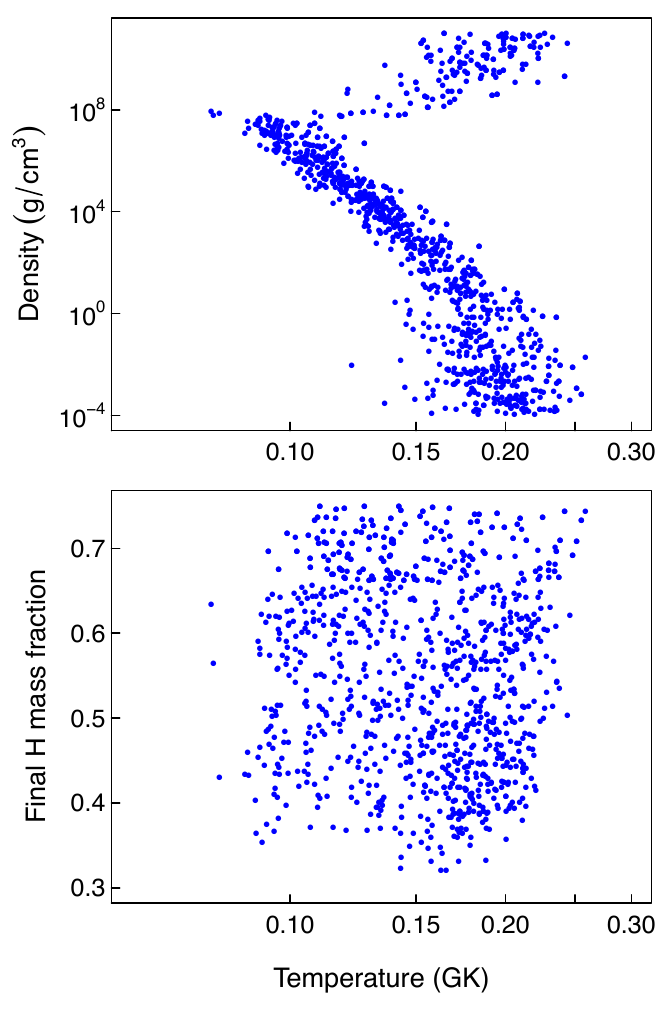}
\caption{(Top) Stellar density versus temperature, and (bottom) final hydrogen mass fraction versus temperature, for sets of ($T$, $\rho$, $X_H^f$) values that reproduce measured elemental abundances in NGC 2419. The results are obtained by random sampling of $T$, $\rho$, $X_H^f$, all nuclear rates in the reaction network, and initial abundances using $10^5$ network samples.\label{fig:MC4}}
\end{figure}

The results of the simultaneous random sampling procedure (for $T$, $\rho$, $X_H^f$, all nuclear reaction rates in the network, and initial composition) are displayed in Figure~\ref{fig:MC4}. In total, $10^5$ reaction network samples were computed. The significant increase in the scatter of the acceptable solutions (blue circles), compared to Figure~\ref{fig:MC2}, is evident. A detailed analysis of which nuclear reaction rate and initial abundance variations have the largest impact on the scatter is beyond the scope of the present work and will be presented in a forthcoming publication. Test calculations showed, for example, that the sampling of the initial $^{20}$Ne and $^{36}$Ar abundances contributes significantly to the observed scatter. 

The temperature and density combinations that provide an acceptable match between simulated and measured abundances are not as well confined in parameter space compared to Figure~\ref{fig:MC2}, but the results of the simultaneous random sampling procedure display similar features and provide important constraints. The simulated abundances of Mg, Si, K, Ca, Sc, Ti, and V simultaneously match the observations for a temperature range of $78$~MK $\le$ $T$ $\le$ $259$~MK (blue circles). Our third main result is that, {\it even if we take the uncertainties in nuclear rates and initial composition into account, we again find a correlation between stellar temperature and density values that provide a satisfactory match between simulated and measured elemental abundances in NGC 2419}.

\section{Polluter candidates}\label{sec:polluters}
Polluter candidates must fulfill a number of necessary conditions. First, their temperatures and densities must give rise to the measured abundance pattern, preferably of all relevant elements. Second, their total ejected matter must account for the observed mass budget. Third, their ejecta must be retained by the globular cluster. In this work, we will focus on the first condition and leave an investigation of the latter conditions, apart from a few general comments, to future work.

\begin{figure}
\epsscale{1.0}
\plotone{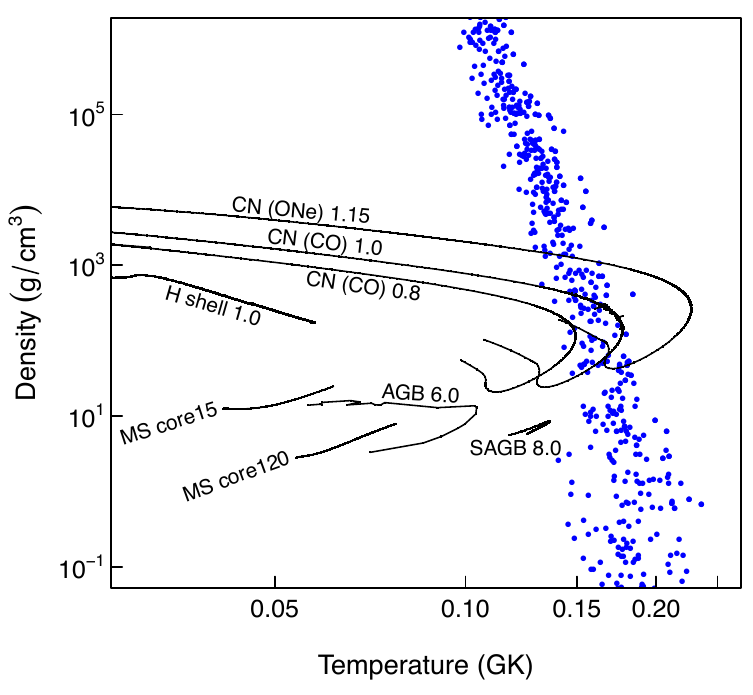}
\caption{Same as the top panel of Figure~\ref{fig:MC4}, but with hydrogen burning $T - \rho$ conditions for several hydrogen-burning polluter candidates superimposed (black solid lines): (i) center of convective hydrogen-burning cores of $15$~M$_\odot$ and $120$~M$_\odot$ stars \citep[``MS core''; from][]{limongi15}; (ii) base of  radiative hydrogen burning shell, from the beginning to the tip of the red giant branch, for a $1$~M$_\odot$ model \citep[``H shell''; from][]{karakas10}; (iii) hot-bottom burning, occurring at the base of the convective hydrogen envelope, during the interpulse period in a $6$~M$_\odot$ thermally-pulsing asymptotic giant branch star \citep[``AGB''; from][]{karakas10}; (iv) hot-bottom burning during the interpulse period in an $8$~M$_\odot$ thermally-pulsing super-asymptotic giant branch star \citep[``SAGB''; from][]{doherty15}; (v) hottest hydrogen-burning zone in classical nova (``CN'') models, involving an underlying carbon-oxygen (CO) or oxygen-neon (ONe) white dwarf \citep[from][]{jose15}. The number after the abbreviation stands for the mass of the stellar model assumed. The metallicities of all models, except for classical novae (see text), are similar to the measured value in NGC 2419 (see Section~\ref{subsec:composition}).
\label{fig:MC5}}
\end{figure}

We show in Figure~\ref{fig:MC5} a magnified section of the top panel of Figure~\ref{fig:MC4}, but add hydrogen burning temperature-density tracks (solid black lines) for several polluter candidates. For two main reasons, we do not expect a potential candidate site to exactly reproduce the $T - \rho$ conditions predicted here. First, the temperature and density, assumed to be constant in our simple model, both vary in realistic hydrogen burning environments. However, these variations are expected to be relatively small during quiescent burning stages. Second, the temperatures predicted here are directly comparable only to radiative, narrow burning regions. For convective regions, on the other hand, the hydrogen fuel burns in a wide zone at an effective temperature, with most of the nucleosynthesis occurring in the hottest zone. However, the difference between the actual temperature in the hottest zone and the effective temperature for the entire region is relatively small \citep[see][]{prantzos07}. In the following, we will adopt our acceptable temperature and density solutions at face value and ask which hydrogen burning environments are able to produce the appropriate conditions.

\subsection{Massive stars}
The two tracks shown in Figure~\ref{fig:MC5} labeled ``MS core 15'' and ``MS core 120'' refer to hydrogen burning in the convective cores of $15$~M$_\odot$ and $120$~M$_\odot$ stars, respectively, and were adopted from the latest models of \citet[][]{limongi15}. The assumed metallicity is [Fe/H] = $-2.0$, which is close to the measured value for NGC 2419 (Section~\ref{subsec:composition}), and the initial rotational speed amounts to $300$~km/s. Both tracks start at the zero-age main sequence and end when the central hydrogen density has fallen to $X_H$ = $0.01$. The maximum temperatures achieved in the $15$~M$_\odot$ and $120$~M$_\odot$ models are $62$~MK and $78$~MK, respectively. The central density at this stage is $24$~g/cm$^3$ and $8$~g/cm$^3$, respectively. The $T - \rho$ values of these models come nowhere near the range of acceptable conditions (blue circles in Figure~\ref{fig:MC5}). The same conclusion holds for hydrogen burning in the cores of supermassive stars (with $M$ $\approx$ $10^4~M_\odot$). Therefore, the abundance anomalies observed in NGC 2419 cannot be produced by any of the scenarios involving hydrogen burning in the cores of massive stars that have been considered in the literature, including rapidly rotating massive stars \citep[][]{decressin07}, massive stars in interacting binary systems \citep[][]{demink09}, or supermassive stars \citep[][]{denissenkov14}.

After hydrogen exhaustion in the core, hydrogen continues to burn in a shell until the burning is turned off by the advancing He burning shell. The physical conditions of the hydrogen burning shell are mainly driven by the underlying core mass. The $15$~M$_\odot$ and $120$~M$_\odot$ models referred to above achieve maximum hydrogen shell temperatures of $84$~MK and $110$~MK, respectively. The corresponding densities are $68$~g/cm$^3$ and $22$~g/cm$^3$, respectively. Again, these conditions (not shown in Figure~\ref{fig:MC5}) are insufficient to account for the measured abundance anomalies in NGC 2419.

\subsection{Low-mass stars}
The hydrogen cores of low-mass stars reach insufficient temperatures to make these sites  viable polluter candidates. For example, the maximum central hydrogen burning temperature is $<$ $30$~MK and $<$ $50$~MK for an $1.0$~M$_\odot$ and a $6.0$~M$_\odot$ model, respectively (not shown in Figure~\ref{fig:MC5}). However, low-mass stars achieve higher temperatures in the hydrogen burning shell. The track labeled ``H shell 1.0'', adopted from \citet[][]{karakas10},  represents the $T - \rho$ conditions for a $1.0$~M$_\odot$ model, with a metallicity of [Fe/H] = $-2.2$, at the base of the radiative hydrogen burning shell, from the beginning to the tip of the red giant branch. The maximum temperature achieved at the end of this track is $57$~MK, when the density amounts to $172$~g/cm$^3$. These values are far smaller compared to the conditions indicated by the blue circles in Figure~\ref{fig:MC5}. This means that the (second generation) low-mass stars measured by \citet[][]{mucciarelli12} and \citet[][]{cohenkirby12} in NGC 2419, all located near the tip of the red giant branch, certainly cannot have produced {\it in situ} the observed abundance anomalies (Section~\ref{sec:intro}). It also means that hydrogen shell burning in first-generation polluter stars cannot account for the reported abundance anomalies. 

\subsection{AGB stars and Super-AGB stars}\label{subsec:agb}
Asymptotic giant branch stars are the evolved descendents of low- and intermediate-mass stars, with masses in the range of $\approx$ $0.8-7$~M$_\odot$, depending on metallicity. They consist of a carbon-oxygen core, surrounded by a helium- and a hydrogen-burning shell, and undergo a series of thermal pulses. The highest hydrogen burning temperatures, and thus the most efficient hydrogen burning nucleosynthesis, in such stars occurs at the bottom of the convective envelope and is referred to as {\it hot-bottom burning}. Stars of higher initial mass reach sufficient temperatures to experience carbon burning in a partially degenerate region near the stellar center and eventually form an oxygen-neon core \citep[][]{ritossa96}. They also ascend the asymptotic giant branch, where they are known as super-asymptotic giant branch (SAGB) stars, undergo a series of thermal pulses, and experience hot-bottom burning. For a recent review see \citet[][]{karakas14}.

The track labeled ``AGB 6.0'', adopted from \citet[][]{karakas10}, shows conditions at the base of the convective hydrogen envelope during the interpulse period (hot-bottom burning), for a $6.0$~M$_\odot$ thermally-pulsing asymptotic giant branch star with a metallicity of [Fe/H] = $-2.2$. The lifetime of this model star is $74$~My. The maximum temperature achieved is about $100$~MK, which is insufficient to reproduce the measured abundance anomalies in NGC 2419. At the densities representative of this track ($\approx$ $10$~g/cm$^3$), the maximum temperature would need to increase significantly, to about $150$~MK, in order to come close to the region occupied by the blue circles. This increase is unlikely, even when fine-turning stellar model parameters such as the mass-loss rate or the mass of the hydrogen-exhausted core prior to the start of the asymptotic giant branch, which determines the maximum hot-bottom burning temperature. 
Therefore, intermediate-mass AGB stars are not favorable candidates for the polluter stars.

We also considered an $8.0$~M$_\odot$, $Z$ = $10^{-4}$ model \citep{doherty15}. The track for hot-bottom burning during the interpulse period of the thermally-pulsing super-asymptotic giant branch phase is labeled ``SAGB 8.0'' in Figure~\ref{fig:MC5}. The lifetime of this model star is $34$~My. The model achieves a maximum hydrogen-burning temperature of $136$~MK at a density of $9$~g/cm$^3$. The models of \citet[][]{ventura13}, which use the Full Spectrum of Turbulence (FST) model of convection, reach a slightly higher maximum temperature of $141$~MK (assuming $7.5$~M$_\odot$ and $Z$ = $3\times10^{-4}$) compared to the present models that adopt the mixing-length theory. However, to get close to the region occupied by the blue cicles, the maximum temperature would need to increase to $150$~MK. Models of super-AGB stars are complex and such an increase in temperature may be achieved by adjusting poorly known model parameters, such as the mass loss rate or the prescription of convective mixing. Therefore, we cannot rule out super-AGB stars as candidate polluters at this time. The parameter space of these stellar models needs to be more fully explored in the future, as advocated by \citet[][]{renzini13} and others.  

\subsection{Novae}
Classical novae involve a white dwarf of carbon-oxygen (CO) or oxygen-neon (ONe) composition accreting hydrogen-rich matter from a main-sequence partner via Roche lobe overflow. The transferred matter carries angular momentum and forms an accretion disk. Subsequently, matter accumulates on the surface of the white dwarf under degenerate conditions. Once explosive conditions are met, a thermonuclear runaway occurs, leading to a violent expulsion of matter \citep[][]{jose06,starrfield08}.

Most published classical nova models have assumed accretion of matter with solar composition, although some models of very low metallicity \citep[$Z$ $\approx$ $10^{-7}$ to $2 \times 10^{-6}$; ][]{jose07} have also been simulated. Since classical nova models accreting matter of a metallicity appropriate for NGC 2419 have not been computed yet, we adopt the models of \citet[][]{jose15} that assume the accretion of {\it solar} metallicity matter from a companion star and a mixing fraction (pre-enrichment) of 50\% between accreted and underlying white dwarf matter prior to the thermonuclear runaway. Figure~\ref{fig:MC5} displays three tracks, labeled ``CN'', for only the hottest hydrogen burning zone during the thermonuclear runaway. Two models involve underlying carbon-oxygen white dwarfs (CO) with masses of $0.8$~M$_\odot$ and $1.0$~M$_\odot$, and one model an oxygen-neon (ONe) white dwarf with a mass of $1.15$~M$_\odot$. It can be seen that during the evolution the tracks for all three models reach the region of the blue circles. Some of the tracks even extend beyond the range of acceptable $T - \rho$ conditions, implying that other zones in these models, that burn hydrogen at lower temperatures and densities, will also eventually reach the region of the blue circles. 

For a metallicity of $Z$ $\approx$ $10^{-4}$ appropriate for NGC 2419, white dwarfs with masses of $M$ $\gtrsim$ $0.8$~$M_\odot$ have a progenitor age of $\lesssim$ $0.5$~Gy, i.e., the time duration from the zero age main sequence to the point of entering the white dwarf cooling curve \citep[][]{romero15}. This leaves sufficient time for novae  involving massive white dwarfs to pollute the intracluster medium before the formation of the second-generation stars\footnote{Let us consider a specific example. For a metallicity of $Z$ = $10^{-4}$, a star with an initial mass of $7$~$M_\odot$ will become a $1.2$~$M_\odot$ white dwarf of O-Ne composition in about $44$~My \citep[][]{doherty15}. The white dwarf needs some time to cool before a nova outburst can take place; if
the white dwarf is initially too luminous, the envelope is not highly degenerate when the thermonuclear runaway develops and only a mild thermonuclear runaway with no mass ejection may occur. Nova simulations have obtained mass ejection for luminosities as high as $L/L_\odot$ = $0.1$ \citep[][]{starrfield85}, $0.3$ \citep[][]{yaron05}, and $1.0$ \citep[][]{hernanz08}. According to \citet[][]{garciaberro97}, it takes only $61$~My for the $1.2$~$M_\odot$ white dwarf of O-Ne composition to cool to a luminosity of $L/L_\odot$ = $0.1$. Therefore, the entire evolution from the zero age main sequence to the point where mass accretion onto the white dwarf can produce a nova outburst takes about $100$~My.}.

Classical novae could thus be interesting polluter candidates, as previously proposed by \citet[][]{smith96}. Recent work also suggested novae involving isolated white dwarfs, i.e., white dwarfs that accrete directly from the intracluster medium, as polluters \citep[][]{maccarone12}. The authors state that a potential problem with their conjecture could be that ``... largely speaking, classical novae do not burn beyond chlorine ...''. On the contrary, once appropriate hydrogen-burning conditions are established (Figure~\ref{fig:MC5}), potassium, for example, will be produced from preexisting argon, as explained in Section~\ref{sec:flows}. 

Novae have so far received little attention in the literature as polluter candidates. Recent work reported a {\it present} frequency of $0.05$ novae per year per globular cluster \citep[][]{henze13}, a value that is much higher than previous estimates for the globular cluster nova rate. The predicted upper limit for the ejected mass per nova is $(2-3)$ $\times$ 10$^{-3}$ M$_\odot$ \citep[][]{shara10}. If we assume $0.5$~Gy for the time period over which novae polluted the intracluster medium before the formation of the second-generation stars (Section~\ref{subsec:general}) and a 10\% efficiency for converting nova ejecta into new stars, we find $\approx$ $0.05$ y$^{-1}$ $\times$ $0.5$ $\times$ $10^9$ y $\times$ $0.1$ $\times$ $10^{-3}$ M$_\odot$ $=$ $2.5 \times 10^3$ M$_\odot$ for the total mass in the cluster that could be processed by novae. On the other hand, NGC 2419 has a mass of $9 \times 10^5$~M$_{\odot}$ (Section~\ref{sec:intro}) and about 30\% of the stars in this cluster are potassium enriched \citep[][]{mucciarelli12,cohenkirby12}, with small dilution factors of $f$ $\le$ $0.04$ (i.e., most of the enriched matter consists of processed rather than pristine material; see Section~\ref{subsec:MC1}). Therefore, the polluters ejected a total mass of $\approx$ $9 \times 10^5$~M$_{\odot}$ $\times$ $0.3$ = $2.7 \times 10^5$ M$_{\odot}$, about two orders of magnitude higher than what we expect from novae. However, some of the above parameters are highly uncertain. For example, the {\it past} nova rate was perhaps much higher, especially if the white dwarfs accreted directly from the dense intracluster medium in the early globular cluster \citep[][]{maccarone12}. The question whether or not novae can quantitatively account for the reported abundance anomalies has to await new detailed models of white dwarfs accreting matter of a composition consistent with NGC 2419, either from a main-sequence companion or directly from the intracluster medium.

\section{Additional constraints from observed He and C, and from future  measurements}\label{sec:future}
\subsection{Helium}\label{subsec:helium}
Photometry of NGC 2419 provides evidence for a spread in the initial helium abundance of the cluster stars. \citet[][]{lee13} inferred that the red giant branch is split into two distinct subpopulations, with 70\% of the stars showing a primordial helium abundance and the other 30\% showing helium mass fractions near 0.42, corresponding to an enhancement of $\Delta Y \approx 0.19$. Furthermore, \citet[][]{dicris15} inferred three distinct populations from the photometry of the horizontal branch: (i) one with an initial helium abundance close to primordial ($Y$ $=$ $0.25$), (ii) a small population with an intermediate-helium abundance of $0.26$ $<$ $Y$ $\lesssim$ $0.29$, and (iii) a large population with a high initial helium abundance of $Y$ $\approx$ $0.36$, which is derived from the extreme blue horizonal branch. 

\begin{figure}
\epsscale{1.0}
\plotone{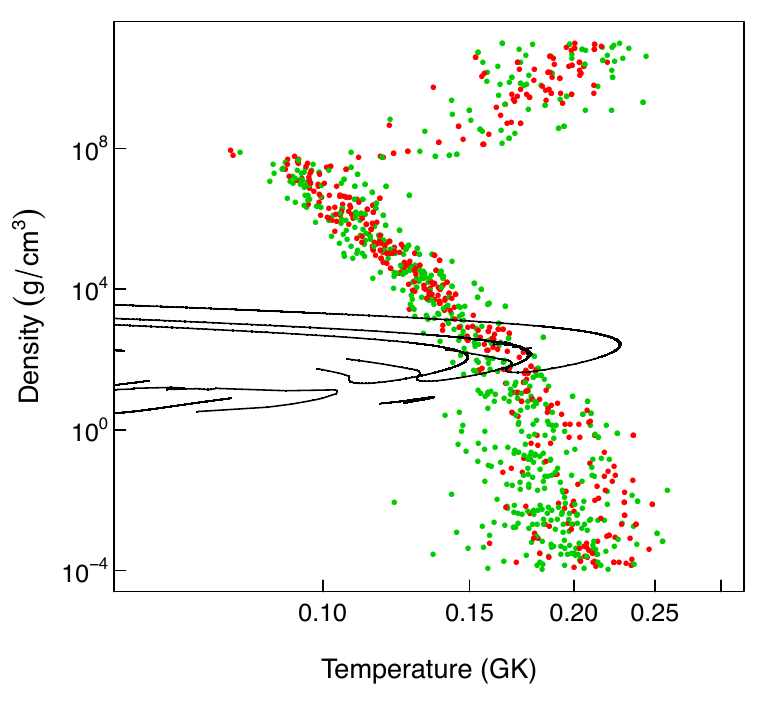}
\caption{Stellar density versus temperature for sets of ($T$, $\rho$, $X_H^f$) values that reproduce measured elemental abundances in NGC 2419. The circles and the $T - \rho$ tracks are the same as those shown in Figure~\ref{fig:MC4} (top) and Figure~\ref{fig:MC5}, respectively. Colors indicate ranges of the final helium mass fraction resulting from our simulations: (red) $0.30$ $\le$ $X_{He}^f$ $\le$ $0.45$; (green) $X_{He}^f$ $>$ $0.45$ or $X_{He}^f$ $<$ $0.30$. The red circles correspond to $T - \rho$ conditions that yield a helium abundance consistent with the recent analysis of extreme populations in NGC 2419 \citep[][]{lee13,dicris15}.
\label{fig:MC7}}
\end{figure}

\citet[][]{dicris15} state that ``...the initial helium abundance of this extreme population is in nice agreement with the predicted helium abundance in the ejecta of massive asymptotic giant branch (AGB) stars of the same metallicity as NGC 2419. This result further supports the hypothesis that second-generation stars in GCs formed from the ashes of intermediate-mass AGB stars...''.  In massive AGB stars and super-AGB stars, most of the surface helium originates from the second dredge-up, with only a minor contribution from hot-bottom burning. In our constant $T - \rho$ model, on the other hand, potassium and helium are concurrently produced during high-temperature hydrogen burning. In Figure~\ref{fig:MC7}, we show the same results as in Figure~\ref{fig:MC4} (top) and in Figure~\ref{fig:MC5}, but now different colors indicate the final helium mass fraction resulting from our simulations. The red circles indicate final helium mass fractions of $0.30$ $\le$ $X_{He}^f$ $\le$ $0.45$, corresponding to the He-rich populations \citep[][]{lee13,dicris15}, while the green circles label solutions with other values of $X_{He}^f$. It can be seen that the elusive polluters that gave rise to the observed Mg-K anticorrelation in NGC 2419 could account simultaneously for the enhanced initial helium abundance (red circles), inferred by \citet[][]{lee13} and \citet[][]{dicris15}, over the entire density range ($\rho$ $=$ $10^{-4} - 10^{11}$~g/cm$^3$) explored in the present work. In other words, polluter candidates other than AGB or super-AGB stars are also able to account for the helium measurements, {\it assuming that K and He are produced during the same high-temperature hydrogen burning process.} If this was indeed the case, then the helium measurements further constrain the hydrogen-burning temperature range of the polluters, as indicated by the reduced scatter of the red circles compared to the green circles in Figure~\ref{fig:MC7}.

\subsection{Lithium}
Our simulated lithium abundance depends strongly on the $T- \rho$ conditions assumed for hydrogen burning and, therefore, lithium measurements would further constrain the parameter space of the polluters. Unfortunately, lithium has not been observed in NGC 2419. According to \citet[][]{cohenkirby12}, ``...the lithium line at $6707$~{\AA} cannot be detected in the summed spectra of either group of NGC 2419 giants...''. Presumably these giant stars have depleted their lithium in the usual way. 

We note that some globular clusters, e.g., NGC 1904 and NGC 2808 \citep[][]{dorazi15}, show a reduced Li content in the proposed second generation stars, with high Na and low O. This is what would be expected from hydrogen burning in the polluters, because Li is destroyed at temperatures as low as $2$~MK. However, there are other globular clusters, e.g., NGC 6397 \citep[][]{lind09}, M 4 \citep[][]{dorazi10,mucciarelli11}, M 12 \citep[][]{dorazi14}, and NGC 362 \citep[][]{dorazi15}, where both generations of stars show essentially the same Li content, which requires that the polluters must also produce Li. This is one reason for the continued investigation of AGB and super-AGB stars as the polluters. However, it is interesting to note that classical novae, implicated by our results, can also produce Li. Of course, both proposed polluter scenarios have quantitative problems in producing the required amount of Li.

\subsection{Carbon, Nitrogen, Oxygen}
Consider again Figure~\ref{fig:abund1}, showing in each panel on the right-hand side the extreme observed stars, consisting of matter that underwent an elusive hydrogen burning process during a previous stellar generation. These extreme stars were likely born with a different He and CNO composition compared to normal stars (located on the left-hand side in each panel of Figure~\ref{fig:abund1}) that were presumably born with a composition similar to field stars of the same metallicity (see Table~\ref{tab:composition}). 

\begin{figure}
\epsscale{1.0}
\plotone{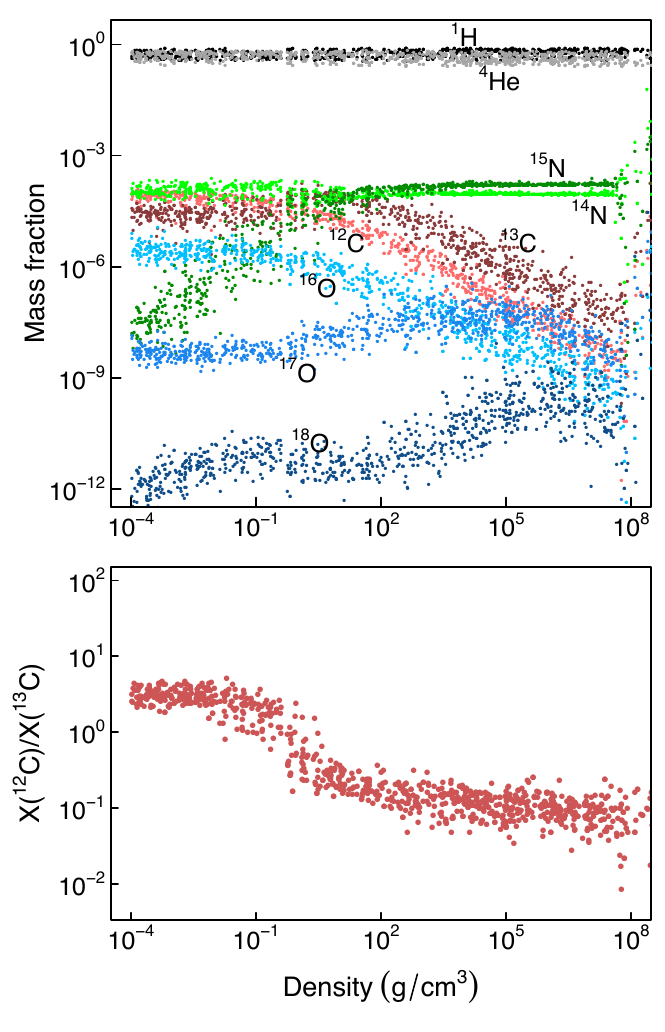}
\caption{Simulated final abundances of $^1$H, $^4$He,$^{12}$C, $^{13}$C, $^{14}$N, $^{15}$N, $^{16}$O, $^{17}$O, and $^{18}$O (top panel) and 
carbon isotopic mass fraction ratio (bottom panel) in processed matter versus stellar density. The results are obtained from the same Monte Carlo simulation shown in Figure~\ref{fig:MC4} (blue circles) and include the random sampling of nuclear reaction rates and initial composition. 
\label{fig:MC6}}
\end{figure}

Figure~\ref{fig:MC6} (top) shows our simulated final abundances of the light nuclides $^1$H, $^4$He,$^{12}$C, $^{13}$C, $^{14}$N, $^{15}$N, $^{16}$O, $^{17}$O, and $^{18}$O in {\it processed} matter. The results are obtained from the same Monte Carlo simulation shown in Figure~\ref{fig:MC4} (blue circles) and include the random sampling of nuclear reaction rates and initial composition. The bottom panel displays the final carbon isotopic ratio. At very low densities ($\rho$ $\lesssim$ $10$~g/cm$^3$), the simulations predict a steadily rising carbon isotopic ratio for decreasing density. At densities above $\rho$ $\approx$ $10^2$~g/cm$^3$, nitrogen is by far the most abundant CNO isotope and the carbon isotopic ratio is $^{12}$C/$^{13}$C $\approx$ $0.1$. 

It would be interesting to compute a number of stellar evolutionary models of low-mass stars, with initial compositions chosen from Figure~\ref{fig:MC6}, and to track the changes in the CNO abundances and in the $^{12}$C/$^{13}$C ratio during the low-mass star evolution to the tip of the red giant branch. The large changes in the initial $^4$He and CNO abundances, compared to canonical models, will have a strong effect in terms of how fast the stars evolve and their location in the color-magnitude diagram; in particular, the helium enrichment will cause the stars to appear hotter and bluer. Therefore, it may be possible to further constrain the $T - \rho$ conditions of the polluters by comparing, for the extreme stars, the measured and the simulated luminosity and elemental carbon abundance. Future measurements of the $^{12}$C/$^{13}$C isotopic ratio could also be important in this regard. This reasoning explicitly assumes that in the polluters the CNO isotopes underwent the same high-temperature hydrogen burning process as the heavier elements (Mg to V), and no additional, non-hydrogen, burning process. We will leave this investigation to future work.

\subsection{Oxygen versus Sodium}\label{subsec:oxna}
\begin{figure}
\epsscale{0.7}
\plotone{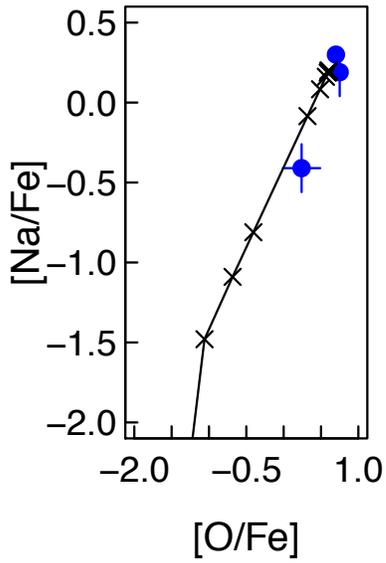}
\caption{Sodium versus oxygen abundance for red giants in NGC 2419 \citep[data from][]{cohenkirby12}. Simulations are shown in black for the conditions $T$ = $160$~MK, $\rho$ = $900$~g/cm$^3$, and $X_H^f$ = $0.70$: the solid line is obtained by mixing one part of processed matter with $f$ parts of pristine matter; the first cross on the left correspond to a dilution factor of $f$ $=$ $0.02$ (mainly processed matter), and the last cross on the right to $f$ $=$ $1000$ (pristine matter). Notice that the simulations predict an O-Na correlation instead of an anticorrelation.\label{fig:abund10}}
\end{figure}

All Galactic globular clusters that have been examined for the O-Na correlation have (so far) shown this signature \citep[see Figure 2 in][]{gratton12}. But as pointed out in Section 2.3, sodium and oxygen have been measured simultaneously in NGC 2419 for only three red giants \citep[][]{cohenkirby12}. The available data, shown in Figure~\ref{fig:abund10}, are not sufficient to establish a relationship between the abundances of these two elements. The solid line shows the results of a simulation with $T$ $=$ $160$~MK, $\rho$ $=$ $900$~g/cm$^3$, and $X_H^f$ $=$ $0.70$, i.e., the same conditions referred to in Figure~\ref{fig:abund1}. The crosses, from left to right, correspond to dilution factors of $f$ $=$ $0.02$ (mostly processed matter) to $f$ $=$ $1000$ (pristine matter). Very similar results are obtained for all $T - \rho$ conditions shown as blue circles in Figure~\ref{fig:MC2}, except at very high densities of $\rho$ $>$ $5 \times 10^7$~g/cm$^3$. 

If oxygen and sodium underwent the same high-temperature hydrogen burning process as the heavier elements (Mg to V), and no additional burning process, then our calculations predict an O-Na {\it correlation}. On the other hand, if instead an O-Na {\it anticorrelation} will be observed, then low-temperature and high-temperature hydrogen burning operated independently in NGC 2419, perhaps in different first-generation stars or at different locations in the same stars (Section~\ref{subsec:strategy}). Future measurements are highly desirable, although the observation of oxygen lines in the cool and metal-poor giants of NGC 2419 will be very challenging.


\section{Summary}\label{sec:summary}
We reported here on the first comprehensive investigation of the parameter space involving temperature, density, consumed hydrogen abundance, thermonuclear reaction rates, and initial chemical composition, to constrain the list of candidate sites that produced the recently measured abundance anomalies in the globular cluster NGC 2419. The observed abundances of magnesium, silicon, and scandium are correlated with potassium, while the abundances of calcium, vanadium, and titanium are nearly constant. These signatures provide important clues regarding their origin. 

We assumed a constant temperature and density model and allow for mixing ({\it dilution}) between nuclearly processed and pristine matter. We investigated under which conditions {\it all of the measured abundances} of elements up to vanadium can be reproduced assuming the full range of dilution factors. Using a reaction network Monte Carlo method, we randomly sampled the stellar temperature, stellar density, and the consumed hydrogen mass fraction, to find conditions that can account for the observations. Variations of thermonuclear reaction rates and initial composition were included in the random sampling. {\it We find a correlation between stellar temperature and density values that provide a satisfactory match between simulated and observed elemental abundances of Mg, Si, K, Ca, Sc, Ti, and V in NGC 2419}. Except at the highest densities ($\rho \gtrsim 10^8$~g/cm$^3$), the acceptable conditions range from $\approx$ $100$~MK at $\approx$ $10^8$~g/cm$^3$ to $\approx$ $200$~MK at $\approx$ $10^{-4}$~g/cm$^3$. 

We reviewed hydrogen burning sites from a nucleosynthesis point of view and find that low-mass stars, AGB stars, massive stars, and supermassive stars are unlikely to account for the measured abundance anomalies in NGC 2419. Super-AGB stars could be viable candidates for the polluter stars if stellar model parameters can be fine-tuned to produce higher temperatures. Novae, either involving CO or ONe white dwarfs, could be interesting polluter candidates, but a current lack of low-metallicity nova models precludes firmer conclusions. Apart from nucleosynthesis considerations, all polluter candidates that have been suggested previously (massive stars, AGB stars, super-AGB stars, interacting binaries, novae, or supermassive stars) have a mass budget problem \citep[see discussion in][]{bastian15}. 

The polluter candidates discussed above and in the literature cover a relevant density range of $10$~g/cm$^3$ $\lesssim$ $\rho$ $\lesssim$ $10^{4}$~g/cm$^3$ (see Figure~\ref{fig:MC5}). As already pointed out, we also find acceptable solutions for matching calculated and measured abundances in NGC 2419 for much higher ($\rho$ $=$ $10^4 - 10^{11}$~g/cm$^3$) and for much lower ($\rho$ $=$ $10^{-4} - 10$~g/cm$^3$) densities, with temperatures in the range of $80$~MK $\lesssim$ $T$ $\lesssim$ $260$~MK. It is not clear at this time which astrophysical environments could give rise to such conditions.

Finally, we discussed the possibility of obtaining additional $T - \rho$ constraints for the (first-generation) polluters by evolving (second-generation) stars with non-canonical initial abundances and by comparing, for the extreme stars in NGC 2419, the model results for the luminosity and elemental carbon abundance to the observations. We also pointed out the importance of new O and Na abundance measurements. If oxygen and sodium underwent the same high-temperature hydrogen burning process as the heavier elements (Mg to V), and no additional burning process, then our simulatuions predict an O-Na {\it correlation} instead of an anticorrelation for NGC 2419. If instead an O-Na {\it anticorrelation} will be observed in the future, then low-temperature and high-temperature hydrogen burning operated independently in NGC 2419, either in different first-generation stars or at different locations in the same stars.

\acknowledgments
We would like to thank Jordi Jos\'e, Marco Limongi, and Alessandro Chieffi for providing up-to-date temperature-density tracks. Helpful comments from Lori Downen, Bart Dunlap, Alex Heger, Fabian Heitsch, Sean Hunt, Richard Longland, Sumner Starrfield, Chris Tout, and David Yong are highly appreciated. AIK is grateful for the support of the NCI NationalÊ Facility at the ANU, and was supported through an Australian Research Council Future Fellowship (FT110100475). This work was supported in part by the U.S. Department of Energy under Contract No. DE-FG02-97ER41041, and under Australian Research CouncilÕs Discovery Projects funding scheme (project number DP120101815). CI was partially supported by the MoCA Distinguished Visitor Program at Monash University.

\clearpage

\begin{deluxetable}{ccccc}
\tabletypesize{\scriptsize}
\tablecaption{Assumed initial ({\it pristine}) composition for present network calculations.\label{tab:composition}}
\tablewidth{0pt}
\tablehead{
\colhead{Nuclide} & \colhead{$X_i$\tablenotemark{a}} & & \colhead{Nuclide} & \colhead{$X_i$\tablenotemark{a}}
}
\startdata
$^{1}$H 		& 7.54E-01 	& & $^{31}$P  		& 1.27E-07 	 \\
$^{2}$H		& 3.93E-05 	& & $^{32}$S  		& 1.64E-05 	  \\
$^{3}$He  		& 2.29E-05 	& & $^{33}$S  		& 1.07E-07 	 \\
$^{4}$He  		& 2.45E-01 	& & $^{34}$S  		& 1.72E-07 	 \\   
$^{6}$Li 		& 6.54E-12 	& & $^{36}$S     	& 3.53E-10	 \\
$^{7}$Li  		& 2.27E-09 	& & $^{35}$Cl 		& 9.58E-08 	 \\
$^{9}$Be  		& 2.21E-12 	& & $^{37}$Cl 		& 1.42E-08 	 \\
$^{10}$B  		& 9.62E-12 	& & $^{36}$Ar 		& 3.79E-06 	 \\
$^{11}$B  		& 4.65E-11 	& & $^{38}$Ar 		& 1.22E-07 	 \\
$^{12}$C  		& 2.98E-05 	& & $^{40}$Ar		& 1.58E-11	 \\
$^{13}$C  		& 1.38E-07 	& & $^{39}$K  		&  {\bf 3.15E-08} \\
$^{14}$N  		& 7.31E-06 	& & $^{40}$K		& 1.39E-10	 \\
$^{15}$N  		& 1.39E-08 	& & $^{41}$K		& 4.65E-09	 \\
$^{16}$O  	& 2.19E-04 	& & $^{40}$Ca		& {\bf 1.15E-06}   \\
$^{17}$O  	& 2.79E-09 	& & $^{42}$Ca         & 7.00E-09	 \\
$^{18}$O  	& 1.41E-08 	& & $^{43}$Ca  	& 1.93E-10	 \\
$^{19}$F 		& 1.22E-09 	& & $^{44}$Ca 		& 1.06E-08	 \\
$^{20}$Ne 	& 8.08E-06 	& & $^{46}$Ca  	& 6.96E-13	 \\
$^{21}$Ne 	& 5.59E-09 	& & $^{48}$Ca 	& 1.94E-08	 \\
$^{22}$Ne 	& 1.32E-07 	& & $^{45}$Sc  	& 3.30E-10	 \\    
$^{23}$Na 	& {\bf 4.30E-07}  & & $^{46}$Ti  	& 8.96E-10	 \\
$^{24}$Mg 	& {\bf 1.53E-05} & & $^{47}$Ti  	& 1.88E-10	 \\
$^{25}$Mg 	& 5.25E-08 	 & & $^{48}$Ti		& 3.92E-08	 \\
$^{26}$Mg 	& 5.81E-08 	& & $^{49}$Ti 		& 9.81E-10	 \\ 
$^{27}$Al 		&  {\bf 2.50E-06} & & $^{50}$Ti 		& 2.08E-10	 \\
$^{28}$Si 		&  {\bf 1.28E-05} & & $^{50}$V 		& 1.13E-12	 \\
$^{29}$Si 		& 1.49E-07 	 & & $^{51}$V  		&  {\bf 2.74E-09} \\
$^{30}$Si 		& 1.15E-07 	 & & & \\    
\enddata
\tablenotetext{a}{Mass fractions adopted from a Galactic chemical evolution model (see text) that reproduces measured abundances in field stars of the same average metallicity as NGC 2419 ([Fe/H] = $-2.1$). Values in bold were adjusted to match observed abundances of red giants in NGC 2419; the original values of the Galactic chemical evolution model were: $X_i$ = 1.14E-07 ($^{23}$Na), 8.26E-06 ($^{24}$Mg), 2.58E-07 ($^{27}$Al), 3.08E-05 ($^{28}$Si), 7.15E-08 ($^{39}$K), 1.85E-06 ($^{40}$Ca), and 1.54E-09 ($^{51}$V). 
}
\end{deluxetable}


\begin{thebibliography}{} 
\bibitem[Andrievsky et al.(2008)]{andrievsky08} Andrievsky, S. M., et al.  2008, \aap, 481, 481
\bibitem[Angelou et al.(2012)]{angelou12} Angelou, G. C., et al.  2012, \apj, 749, 128
\bibitem[Bastian et al.(2013)]{bastian13} Bastian, N., et al.  2013, \mnras, 436, 2398
\bibitem[Bastian, Cabrera-Ziri \& Salaris(2015)]{bastian15} Bastian, N., Cabrera-Ziri, I., \& Salaris, M.  2015, \mnras, 449, 3333
\bibitem[Bellazzini et al.(2013)]{bellazzini13} Bellazzini, M., et al.  2013, Mem. S.A.It., 84, 175
\bibitem[Carretta et al.(2009)]{carretta09} Carretta E., Bragaglia A., Gratton R., \& Lucatello S.  2009, \aap,
505, 139
\bibitem[Cohen \& Kirby(2012)]{cohenkirby12} Cohen, J. G., \& Kirby, E. N.  2012, \apj, 760, 86
\bibitem[D'Antona et al.(2002)]{dantona02} D'Antona, F. et al.  2002, \aap, 395, 69  
\bibitem[Decressin et al.(2007)]{decressin07} Decressin, T. et al.  2007, \aap, 464, 1029
\bibitem[Decressin, Baumgardt \& Kroupa(2008)]{decressin08} Decressin, T., Baumgardt, H., \& Kroupa, P.  2008, \aap, 492,101 
\bibitem[de Mink et al.(2009)]{demink09} de Mink, S. E., Pols, O. R., Langer, N., \& Izzard, R. G.  2009, \aap, 507, L1
\bibitem[Denissenkov \& Hartwick(2014)]{denissenkov14} Denissenkov, P. A., \& Hartwick, F. D. A.  2014, \mnras, 437, L21
\bibitem[D'Ercole et al.(2008)]{dercole08} D'Ercole, A. et al.  2008, \mnras, 391, 825
\bibitem[Di Criscienzo et al.(2011)]{dicris11} Di Criscienzo, M. et al.  2011, \mnras, 414, 3381
\bibitem[Di Criscienzo et al.(2015)]{dicris15} Di Criscienzo, M. et al.  2015, \mnras, 446, 1469
\bibitem[Doherty et al.(2014)]{doherty14} Doherty, C. L., et al.  2014, \mnras, 441, 582 
\bibitem[Doherty et al.(2015)]{doherty15} Doherty, C. L., et al.  2015, \mnras, 446, 2599 
\bibitem[D'Orazi \& Marino(2010)]{dorazi10} D'Orazi, V., \& Marino, A.F.  2010, \apj, 716, L166
\bibitem[D'Orazi et al.(2014)]{dorazi14} D'Orazi, V., et al.  2014, \apj, 791, 39
\bibitem[D'Orazi et al.(2015)]{dorazi15} D'Orazi, V., et al.  2015, \mnras, 449, 4038
\bibitem[Forbes \& Bridges(2010)]{forbes10} Forbes, D.A., \& Bridges, T.  2010, \mnras, 404, 1203
\bibitem[Fuller, Fowler \& Newman(1980)]{fuller80} Fuller, G. M., Fowler, W. A., \& Newman, M. J.  1980, \apjs, 42, 447
\bibitem[Fuller, Fowler \& Newman(1982)]{fuller82} Fuller, G. M., Fowler, W. A., \& Newman, M. J.  1982, \apjs, 48, 279
\bibitem[Garc\'ia-Berro, Isern \& Hernanz(1997)]{garciaberro97} Garc\'ia-Berro, E., Isern, J., \& Hernanz, M.  1997, \mnras, 289, 973
\bibitem[Goswami \& Prantzos(2000)]{goswami00} Goswami, A., \& Prantzos, N.  2000, \aap, 359, 191
\bibitem[Gratton et al.(2000)]{gratton00} Gratton, R.G. et al.  2000, \aap, 354, 169
\bibitem[Gratton et al.(2001)]{gratton01} Gratton, R.G., et al.  2001, \aap, 369, 87
\bibitem[Gratton, Sneden \& Carretta(2004)]{gratton04} Gratton, R., Sneden, C. \& Carretta, E.  2004, \araa, 42, 385
\bibitem[Gratton, Carretta \& Bragaglia(2012)]{gratton12} Gratton, R. G., Carretta, E. \& Bragaglia, A.  2012, Astron. Astrophys. Rev., 20, 50
\bibitem[Henze et al.(2013)]{henze13} Henze, M., et al.  2013, \aap, 549, A120
\bibitem[Hernanz \& Jos\'e(2008)]{hernanz08} Hernanz, M., \& Jos\'e, J. 2008, New Astron. Rev., 52, 386
\bibitem[Ibata et al.(2011a)]{ibata11a} Ibata, R. et al.  2011, \apj, 738, 186
\bibitem[Ibata et al.(2011b)]{ibata11b} Ibata, R. et al.  2011, \apj, 743, 43
\bibitem[Iliadis et al.(2011)]{iliadis11} Iliadis, C. et al.  2011, \apjs, 193, 16
\bibitem[Iliadis(2015)]{iliadis15} Iliadis, C. 2015, {\it Nuclear Physics of Stars} (2nd ed.; Weinheim: Wiley-VCH)
\bibitem[Iliadis et al.(2015)]{iliadis15b} Iliadis, C., et al.  2015, J. Phys. G: Nucl. Part. Phys., 42, 034007
\bibitem[James et al.(2004)]{james04} James, G. et al.  2004, \aap, 427, 825
\bibitem[Jos\'e, Hernanz \& Iliadis(2006)]{jose06} Jos\'e, J., Hernanz, M., \& Iliadis, C.  2006, Nucl. Phys. A, 777, 550
\bibitem[Jos\'e et al.(2007)]{jose07} Jos\'e, J., Garc\'ia-Berro, E., Hernanz, M., \& Gil-Pons, P.  2007, \apjl, 662, 103
\bibitem[Jos\'e (2015)]{jose15} Jos\'e, J.  2015 (private communication)
\bibitem[Karakas(2010)]{karakas10} Karakas, A. I.  2010, \mnras, 403, 1413
\bibitem[Karakas \& Lattanzio(2014)]{karakas14} Karakas, A.I., \& Lattanzio, J.C.  2014, \pasa, 31, e030
\bibitem[Kobayashi, Karakas \& Umeda(2011)]{kobayashi11} Kobayashi, C., Karakas, A. I., \& Umeda, H.  2011, \mnras, 414, 3231
\bibitem[Lee at al.(2013)]{lee13} Lee, Y.-W., et al.  2013, \apj, 778, L13
\bibitem[Limongi \& Chieffi(2015)]{limongi15} Limongi, M., \& Chieffi, A.  2015 (private communication)
\bibitem[Lind et al.(2009)]{lind09} Lind, K., et al.  2015, \aap, 503, 545
\bibitem[Lind et al.(2015)]{lind15} Lind, K., et al.  2015, \aap, 575, L12
\bibitem[Longland et al.(2010)]{longland10} Longland, R. et al.  2010, Nucl. Phys. A, 841, 1
\bibitem[Maccarone \& Zurek(2012)]{maccarone12} Maccarone T. J., \& Zurek D. R.  2012, \mnras, 423, 2
\bibitem[Mackey \& van den Bergh(2005)]{mackey05} Mackey, A. D., \& van den Bergh, S.  2005, \mnras, 360, 631
\bibitem[Meszaros et al.(2015)]{meszaros15} Meszaros S. et al.,  2015, ArXiv e-prints
\bibitem[Moore \& Bildsten(2011)]{moore11} Moore, K., \& Bildsten, L.  2011, \apj, 728, 81
\bibitem[Mucciarelli et al.(2011)]{mucciarelli11} Mucciarelli, A. et al.  2011, \mnras, 412, 81
\bibitem[Mucciarelli et al.(2012)]{mucciarelli12} Mucciarelli, A. et al.  2012, \mnras, 426, 2889
\bibitem[Mucciarelli et al.(2015)]{mucciarelli15} Mucciarelli, A. et al.  2015, \apj, 801, 68
\bibitem[Oda et al.(1994)]{oda94} Oda, T. et al.  1994, Atom. Data Nucl. Data Tab., 56, 231
\bibitem[Piotto et al.(2007)]{piotto07} Piotto, G. et al.  2007, \apj, 661, L53
\bibitem[Powell et al.(1999)]{powell99} Powell, D. C. et al.  1999, Nucl. Phys. A, 660, 349
\bibitem[Prantzos \& Charbonnel(2006)]{prantzos06} Prantzos, N., \& Charbonnel, C.  2006, \aap, 458, 135
\bibitem[Prantzos, Charbonnel \& Iliadis(2007)]{prantzos07} Prantzos, N., Charbonnel, C., \& Iliadis, C.  2007, \aap, 470, 179
\bibitem[Renzini(2013)]{renzini13} Renzini, A.  2013, Mem. S.A.It., 84, 162
\bibitem[Ritossa et al.(1996)]{ritossa96} Ritossa, C, Garc\'ia-Berro, E., \& Iben, I.  1996, \apj, 460, 489
\bibitem[Romero, Campos \& Kepler(2015)]{romero15} Romero, A.D., Campos, F., \& Kepler, S.O.  2015, \mnras, 450, 3708
\bibitem[Sallaska et al.(2013)]{sallaska13} Sallaska, A. et al.  2013, \apjs, 207, 18
\bibitem[Shara et al.(2004)]{shara04} Shara, M.M. et al.  2004, \apj, 605, L117
\bibitem[Shara et al.(2010)]{shara10} Shara, M.M. et al.  2010, \apj, 725, 831
\bibitem[Siess(2010)]{siess10} Siess, L.  2010, \aap, 512, A10
\bibitem[Sills \& Glebbeek(2010)]{sills10} Sills, A., \& Glebbeek, E.  2010, \mnras, 407, 277
\bibitem[Smith \& Kraft(1996)]{smith96} Smith, G.H., \& Kraft, R.P.  1996, \pasp, 108, 344
\bibitem[Starrfield, Sparks \& Truran(1985)]{starrfield85} Starrfield, S., Sparks, W.M., \& Truran, J.W.  1985, \apj, 291, 136
\bibitem[Starrfield, Iliadis \& Hix(2008)]{starrfield08} Starrfield, S., Iliadis, C., \& Hix, W. R.  2008, in {\it Classical Novae}, ed. M. F. Bode \& A. Evans (2nd ed.; Cambridge: Cambridge Univ. Press), 77
\bibitem[Villanova et al.(2007)]{villanova07} Villanova, S. et al.  2007, \apj, 663, 296
\bibitem[Ventura et al.(2001)]{ventura01} Ventura, P., D'Antona, F., Mazzitelli, L., \& Gratton, R., 2001  \apj, 550, L65
\bibitem[Ventura et al.(2012)]{ventura12} Ventura, P. et al., 2012  \apj, 761, L30
\bibitem[Ventura et al.(2013)]{ventura13} Ventura, P., Di Criscienzo, M., Carini, R., and D'Antona, F.  2013, \mnras, 431, 3642
\bibitem[Yaron et al.(2005)]{yaron05} Yaron, O., Prialnik, D., Shara, M.M., \& Kovetz, A.  2005, \apj, 623, 398
\end{thebibliography}
\end{document}